\documentclass[a4paper,12pt]{article}
\pdfoutput=1

\usepackage{amssymb,amsmath,bm}
\usepackage{a4wide}
\usepackage{color}
\usepackage{slashed}
\usepackage{graphicx}
\usepackage{amsfonts}
\usepackage{lscape}
\usepackage{amsthm}
\usepackage{booktabs}
\usepackage{multirow,geometry,mathrsfs}
\usepackage{array}
\usepackage{cancel}
\usepackage{rotating}
\usepackage[numbers, sort]{natbib}
\usepackage{float}
\usepackage{relsize,exscale}
\usepackage[toc,page]{appendix}
\usepackage{multicol}
\usepackage{color}
\usepackage{comment}
\usepackage{bbm}
\usepackage{calligra}
\usepackage{calrsfs}
\definecolor{azur}{rgb}{0,0.498,1}
\definecolor{caeruleum}{rgb}{0.208,0.478,0.718}
\definecolor{indigo_elec}{rgb}{0.435,0,1}

\DeclareMathAlphabet{\mathcalligra}{T1}{calligra}{m}{n}

\makeatletter
\def\clap#1{\hbox to 0pt{\hss #1\hss}}%
\def\ligne#1{%
  \hbox to \hsize{%
    \vbox{\centering #1}}}%
\def\haut#1#2#3{%
  \hbox to \hsize{%
    \rlap{\vtop{\raggedright #1}}%
    \hss
    \clap{\vtop{\centering #2}}%
    \hss
    \llap{\vtop{\raggedleft #3}}}}%
\def\bas#1#2#3{%
  \hbox to \hsize{%
    \rlap{\vbox{\raggedright #1}}%
    \hss
    \clap{\vbox{\centering #2}}%
    \hss
    \llap{\vbox{\raggedleft #3}}}}%
\def\maketitle{%
  \thispagestyle{empty}\vbox to \vsize{%
    \haut{}{\Large \@blurb}{}
    \vfill
    \vfill
    \ligne{\Huge \@title}
    \vspace{5mm}
    \ligne{\Large \@author}
    \vspace{1cm}
    \vfill
    \vfill
    \bas{}{\@location \@date}{}
    }%
  \cleardoublepage
  }
\def\date#1{\def\@date{#1}}
\def\author#1{\def\@author{#1}}
\def\title#1{\def\@title{#1}}
\def\location#1{\def\@location{#1}}
\def\blurb#1{\def\@blurb{#1}}
\date{\today}
\author{}
\title{}
\location{}
\blurb{}

\renewenvironment{abstract}{
\null\vfil
\@beginparpenalty\@lowpenalty
\begin{center}
\bfseries \abstractname
\@endparpenalty\@M
\end{center}
}

\makeatother

\usepackage{eurosym}
\usepackage{feynmp}
\usepackage[small,bf]{caption}
\newcommand{\be}{\begin{equation}}
\newcommand{\ee}{\end{equation}}

\newcommand{\cosb}{c_\beta}
\newcommand{\sinb}{s_\beta}
\newcommand{\sbt}{s_{2\beta}}
\newcommand{\cbt}{c_{2\beta}}

\def\non{\nonumber}
\def\noi{\noindent}
\newcommand{\lv}{\Lambda_v}

\newcommand{\beqn}{\begin{eqnarray}}
\newcommand{\eeqn}{\end{eqnarray}}
\newcommand{\bea}{\begin{eqnarray}}
\newcommand{\ena}{\end{eqnarray}}
\newcommand{\ra}{\rightarrow}
\def\neuto{\tilde\chi^0_1}

\def\neutt{\tilde\chi^0_2}
\def\neuth{\tilde\chi^0_3}
\def\neutf{\tilde\chi^0_4}
\def\neutfi{\tilde\chi^0_5}

\def\charg{\tilde\chi^+_1}
\def\chargt{\tilde\chi^+_2}
\def\chargm{\tilde\chi^-_1}

\def\drbar{\overline {\textrm{DR}}}

\def\tb{t_\beta}
\def\ppi{\mathbbm{p}_i}
\def\l{\lambda}

\def\d{\delta}

\def\b{\beta}

\def\k{\kappa}

\numberwithin{equation}{section}

\begin{document}

\begin{titlepage}
\begin{center}
\vspace*{-1cm}
\begin{flushright}
LAPTH-011/17\\
LPT-Orsay-17-14\\
\end{flushright}

\vspace*{1.6cm}
{\Large\bf One-loop renormalisation of the NMSSM in SloopS \\ 2:  the Higgs sector} 

\vspace*{1cm}\renewcommand{\thefootnote}{\fnsymbol{footnote}}

{\large 
G.~B\'elanger$^{1}$\footnote[1]{Email:belanger@lapth.cnrs.fr},
V.~Bizouard$^{1}$\footnote[2]{Email: vincent.bizouard2@gmail.com}
F.~Boudjema$^{1}$\footnote[3]{Email: boudjema@lapth.cnrs.fr },
G.~Chalons$^{2}$\footnote[4]{Email: guillaume.chalons@th.u-psud.fr},
}

\vspace*{0.3cm}

\Large{\today}

\renewcommand{\thefootnote}{\arabic{footnote}}

\vspace*{1cm} 
{\normalsize \it 
$^1\,$LAPTh, Universit\'e Savoie Mont Blanc, CNRS, B.P.110, F-74941 Annecy-le-Vieux Cedex, France\\[2mm]
$^2\,${LPT (UMR8627), CNRS, Univ. Paris-Sud, Universit\'{e} Paris-Saclay, 91045
Orsay, France\\[2mm]
}}

\vspace{0.3cm}

\end{center}
\begin{abstract}
We present a full one-loop renormalisation of the Higgs sector of the Next-to-Minimal-Supersymmetric-Standard-Model (NMSSM) and its
implementation within {\tt SloopS}, a code for the automated computations of
one-loop processes in theories beyond the Standard Model. The present work is the sequel to the study we performed 
on the renormalisation of the sectors of the NMSSM comprising neutralinos, charginos and sfermions thereby completing the full one-loop renormalisation of the NMSSM. 
We have investigated several renormalisation schemes based on alternative choices (on-shell or $\drbar$) of the
physical parameters. Special attention is paid to the issue of the mixing between physical fields. To weigh the impact of the different renormalisation schemes, the  
partial widths for the decays of  the  Higgs bosons into supersymmetric particles are computed at one-loop. In many decays large differences between the schemes are found. We discuss the origin of these differences. In particular we study two contrasting scenarios. The first model is  MSSM-like with a small value for the mixing between the doublet and singlet Higgs superfields while the second model  has a moderate value for this mixing.  We critically discuss the issue of the reconstruction of the underlying parameters and their counterterms in the case of a theory with a large number of parameters, such as the NMSSM, from a set of physical parameters. In the present study this set corresponds to the minimum set of masses for the implementation of the on-shell schemes. 

\end{abstract}

\end{titlepage}

\section{Introduction}
\label{sec:intro}
The discovery of a Standard Model-like Higgs boson at the LHC~\cite{Aad:2012tfa,Chatrchyan:2012xdj} has raised some concerns with one of the favourite extensions of the Standard Model, SM, the minimal supersymmetric standard model (MSSM).
A Higgs mass of 125 GeV, near the maximum value achievable in this model,
requires some fine-tuning~\cite{Hall:2011aa}. In a minimal singlet extension of
the model with a $\mathbb{Z}_3$ symmetry, the Next-to-MSSM or NMSSM, additional
Higgs quartic
couplings allow to raise the tree-level mass of the SM-like
Higgs~\cite{Ellwanger:2009dp,Ellwanger:2006rm}  hence reducing both the amount of
radiative corrections required  from the top/stop sector and the amount of 
fine-tuning~\cite{Ellwanger:2011mu,BasteroGil:2000bw,Ross:2012nr}.
In addition, this model  provides a natural explanation for the scale of the
higgsino parameter $\mu$, by relating it to the vacuum expectation value,  or vev, of the scalar singlet, thus
solving the little hierarchy problem of the MSSM. \\

\noi The computation of higher-order corrections to Higgs production  at the LHC
as well as its  decay rates  has been a field of intense activity for the last
two decades, in particular for the SM Higgs,  for an update
see~\cite{Dittmaier:2011ti,Dittmaier:2012vm,Heinemeyer:2013tqa,
deFlorian:2016spz}. The bulk of the corrections having to do with QCD
corrections. For the MSSM, the best example for the  importance of the higher
order corrections in the Higgs sector is the correction to the Higgs
mass~\cite{Haber:1990aw,Ellis:1990nz}. Without these corrections driven by the
top mass, the MSSM would not have survived for so
long~\cite{Djouadi:2005gj,Carena:2002es,Martin:2004kr,Heinemeyer:2004ms}. Many
of these computations have been extended and/or adapted to the case of the
NMSSM, in particular for the Higgs
masses~\cite{Ellwanger:1993hn,Elliott:1993uc,Elliott:1993bs,Pandita:1993hx,
Pandita:1993tg,Ellwanger:2005fh,Ender:2011qh} with improvements including 
several two-loop
effects~\cite{Degrassi:2009yq,Goodsell:2014pla,Staub:2015aea,Drechsel:2016jdg,
Drechsel:2016htw}. Adaptation of the computations of higher order corrections for Higgs production at the LHC from the SM and the MSSM to the NMSSM have  been performed~\cite{Liebler:2015bka}. One-loop  corrections to Higgs decays in the NMSSM have also been considered with varying degree of generalisation and approximation depending on the final state. Full QCD/SUSY-QCD corrections to decays to SM fermions~\cite{Baglio:2013vya} have been performed, as well as electroweak and QCD corrections to channels such as the decays of CP-odd Higgses into stops~\cite{Baglio:2015noa} and Higgs self-couplings~\cite{Nhung:2013lpa}, while many other channels like decays to neutralinos and charginos have been adapted from the MSSM~\cite{Baglio:2013iia}. Some electroweak corrections are still not fully systematically included for all decays. Many of these one-loop (or in the case of masses beyond one-loop corrections) have been incorporated in several public codes for the NMSSM: {\tt NMSSMTools}~\cite{Ellwanger:2005dv,Ellwanger:2006rn},
{\tt SPheno}~\cite{Staub:2010ty,Porod:2011nf}, {\tt NMSDECAY}~\cite{Das:2011dg}, {\tt SoftSUSY}~\cite{Allanach:2013kza},  
{\tt NMSSMCALC}~\cite{Baglio:2013iia} and {\tt FlexibleSUSY}~\cite{Athron:2014yba}. Most of these computations are based on a $\drbar$ scheme or on a mixed $\drbar$/On-Shell (OS)  scheme as in {\tt NMSSMCALC}~\cite{Baglio:2013iia}. In principle an automated implementation of  two-body decays in $\drbar$ of the NMSSM could be attempted for the NMSSM along the lines described in~\cite{Goodsell:2017pdq}. A full OS scheme at one-loop for the NMSSM has not been studied. \\

\noi One of the aims of this paper is to precisely implement different renormalisation schemes including a few variants of a full OS scheme in order to perform  complete one-loop corrections for \underline{any} process in the NMSSM. We have shown in a previous paper~\cite{Belanger:2016tqb}
 how such a programme is applied to the renormalisation of the
neutralino/chargino and sfermion sectors of the NMSSM, we extend it here to
cover the Higgs sector. Because of  the r\^ole played by the effective $\mu$
parameter in the NMSSM, or in other words, the doublet-singlet $\lambda$ mixing,
there is a strong interconnection between the chargino/neutralino sector and the
Higgs sector which warrants a common and overall coherent approach to the
complete one-loop renormalisation of the NMSSM. 
This work is a natural extension of the work done for the MSSM in
~\cite{Baro:2008bg,Baro:2009gn}  where, after performing the complete
renormalisation of the MSSM, one-loop corrections to masses, two-body decays and
production cross sections at colliders were computed together with one-loop
corrections for various dark matter annihilation
processes~\cite{Baro:2007em, Baro:2009ip,Baro:2009na}. The fact that one is able
to perform one-loop corrections to a host of processes is made possible by the
implementation of the theoretical set-up for the one-loop renormalisation in
{\tt SloopS}, a code for the automated generation and evaluation of any cross
section. The one-loop theoretical set-up will be detailed here. As a
pre-requisite for {\tt SloopS}, one first needs to read a model-file. 
The model file, the NMSSM in this case,  is  obtained automatically with an improved version of    LanHEP~\cite{Semenov:1998eb,Semenov:2008jy,Semenov:2014rea} that allows for   the  generation of the counterterms and the corresponding Feynman rules. The code
then relies on   {\tt FeynArts}~\cite{Hahn:2000kx}, {\tt FormCalC}~\cite{Hahn:1998yk} and {\tt LoopTools} for the automatic computation of one-loop processes~\cite{Hahn:2000jm} including both electroweak and QCD corrections. Preliminary applications of  {\tt SloopS} to  the NMSSM dealt with computing one-loop induced decays into photons: {\it i)} neutralino annihilations into photons, the  gamma-ray lines  for
Dark Matter indirect detection \cite{Chalons:2011ia,Chalons:2012xf}, {\it ii)} Higgs
decays to photons at the LHC \cite{Chalons:2012qe,Belanger:2014roa}. These processes do not  call for counterterms or renormalisation at one-loop yet an important part of the machinery of {\tt SloopS} is called for.  \\

\noi The NMSSM is a typical beyond the SM theory with many
parameters, fields and mixings  where the different sectors are intertwined. The
vast majority of its parameters, as they appear at the level of the Lagrangian,  are not directly measurable in experiments in the sense that there is not a straightforward linear mapping between these parameters and an observable such as a mass. The reconstruction of these parameters is a real challenge even when attempted at tree-level. Renormalisation being tightly linked to the choice of input parameters to be
extracted from experimental measurements, having currently no sign of
supersymmetry leaves this choice with no clear guidance. However, the
current extensive program of precision measurements of the Higgs couplings at
the LHC requires nonetheless precise theoretical  predictions  making the
renormalisation of the Higgs sector of the NMSSM highly desirable. As for the
neutralino/chargino sector that we studied in ~\cite{Belanger:2016tqb}, different schemes are
possible. We use mostly on-shell schemes where input parameters are taken from
the masses of the neutralino/chargino and from the Higgs sectors. In such schemes, based on
two-points functions only, the task of renormalising the model boils down to
choosing a minimal/sufficient set of physical masses as input parameters. In this
work, we have adopted several  sets of input parameters and
discussed how efficiently each set can constrain the needed counterterms to keep
the radiative corrections under control. Moreover, the renormalisation procedure
induces additional mixing, not only among the Higgs physical states but also new
gauge-Higgs and Goldstone-Higgs transitions in the pseudoscalar and charged
sector appear. Such mixing must vanish for on-shell physical states by imposing
appropriate conditions on the wave function renormalisation constants which have
to satisfy Ward-Slavnov-Taylor (WST) identities. In doing so we have rederived
the WST identities governing the $A_i^0 Z^0/H^\pm W^\pm$ mixing in the NMSSM. \\

\noi The dependence on the renormalisation scheme is then illustrated in numerical
computations of  observables. Full one-loop
electroweak  corrections to decays of Higgs
particles are computed. The scheme dependence for Higgs to Higgs decays and for decays
involving charginos/neutralinos is carefully examined.  Note that while the Higgs mass
computation is not as accurate as in other codes (only one-loop corrections are
included), we nevertheless stress that our approach allows for  a consistent
treatment of on-shell renormalisation and one-loop corrections to masses, decays
and scattering processes. This is of importance given the very precise
experimental measurements achieved both for the Higgs and for Dark Matter observables. \\

\noi The paper is organised as follows. Section~\ref{sec:higgs_sector} contains a description of the Higgs sector of the
NMSSM and enumerates the number of fields and parameters which will need
renormalisation. The needed counterterms to obtain ultraviolet finite results
are introduced in Section~\ref{sec:ren_higgs} and in the following section the issue of mixing in
the Higgs sector  through the self-energies is
discussed. Section~\ref{sec:ren_schemes} presents the different renormalisation
schemes which enable a reconstruction of the counterterms of the underlying
parameters the NMSSM with a special attention to those of the Higgs sector. Section~\ref{sec:beta_dep} presents how the numerical results
are checked and how the scheme dependence can be quantified in order to gain insights on
the theoretical uncertainties.  In section~\ref{sec:benchmark} we present
two scenarios for which, in Section~\ref{sec:higgs_decays}, we compute numerically several Higgs
partial widths and discuss their scheme dependence. Finally our conclusions are made in Section~\ref{sec:conclusions}.

\section{The Higgs sector of the NMSSM}
\label{sec:higgs_sector}
\subsection{Fields and potential}

The NMSSM contains three Higgs superfields : two $SU(2)_L$ doublets $\hat{H}_u$ and $\hat{H}_d$, as in the MSSM, and one additional gauge singlet $\hat{S}$
\begin{equation}
\hat{H}_u=\begin{pmatrix}
\hat{H}_u^+\\
\hat{H}_u^0
\end{pmatrix},\quad \hat{H}_d=\begin{pmatrix}
\hat{H}_d^0\\
\hat{H}_d^-
\end{pmatrix},\quad \hat{S}.
\end{equation}
In the $\mathbb{Z}_3$ implementation that we will assume, the Higgs
superpotential is made up of two operators, associated with the dimensionless
couplings $\lambda$ and $\kappa$,
\begin{equation}
W_{Higgs}=-\lambda\hat{S}\hat{H}_d\cdot \hat{H}_u+\frac{1}{3}\kappa\hat{S}^3,
\end{equation}
where $\hat{H}_d \cdot\hat{H}_u = \epsilon_{ab} \hat{H}_d^a \hat{H}_u^b$ and $\epsilon_{ab}$ is the two dimensional
Levi-Civita symbol with $\epsilon_{12} = 1$. The two parameters $\lambda$ and
$\kappa$ of the superpotential will, by construction, affect the phenomenology
of both the Higgs and chargino/neutralino sectors. The five  neutralinos of the
NMSSM will be an admixture of the {\it i)}$SU(2)$, $\tilde{w}$, and $U(1)$,
$\tilde{b}$, neutral gauginos, {\it ii)} the two neutral higgsinos,
$\tilde{h}_{u,d}$, the fermionic components of  two superfields $\hat{H}_{u,d}$
and {\it iii)} the singlino, $\tilde{s}$, the fermionic component of $\hat{S}$. 
$\lambda$ in particular is
crucial, not only it is necessary in order to induce the $\mu$ term but it also gives
rise to mixing in the neutralino sector as well as in the Higgs sector between
the Higgs doublets and the new singlet. In passing we recall that $\mu$ sets the mass scale for the higgsinos,  see \cite{Belanger:2016tqb} for more detail. For the purpose of parameter counting
and of the renormalisation of the Higgs sector at one-loop there is no need to
go over the Yukawa superpotential which we have given in the previous paper
\cite{Belanger:2016tqb}. However we do need to clearly specify again the soft
SUSY breaking Lagrangian, in particular the part relating to the Higgs sector.
\begin{align}
-\mathcal{L}_{soft,scalar}&=m_{H_u}^2|H_u|^2+m_{H_d}^2|H_d|^2+m_S^2|S|^2 \notag\\
&  \quad \quad \quad + (\lambda A_{\lambda}H_u\cdot H_dS+\frac{1}{3}\kappa A_{\kappa}S^3+h.c)
\label{softLagr-scalar}
\end{align}
The first two terms in the first line represent the soft mass terms for the
Higgs doublets and the third, not present in the MSSM, of the
singlet. The second line, not present in the MSSM also, represents the
NMSSM trilinear Higgs couplings $A_\kappa, A_\lambda$.  $A_\lambda$ affects the
mixing between the Higgs doublets and the singlet, beside the mixing introduced
by $\lambda$. This parameter plays an important role in the phenomenology of the
Higgs sector in the NMSSM, note that it gives rise to a Higgs tri-linear
coupling $H_u H_d S$.  No source of CP violation is assumed. 

We are now in a position to write the Higgs potential whose parameters will need to be renormalised. With $g,g^\prime$ being respectively the $SU(2)$ weak and $U(1)$ hypercharge gauge couplings and specifying the components of the doublets, 
$$H_d=\begin{pmatrix}H_d^0\\H_d^-\end{pmatrix} \;\;\;\;
H_u=\begin{pmatrix}H_u^+\\H_u^0\end{pmatrix} $$ the potential reads

\begin{align}
V_{Higgs}=&|\lambda(H_u^+H_d^--H_u^0H_d^0)+\kappa S^2|^2+(m_{H_u}^2+|\lambda S|^2)\left(|H_u^0|^2+|H_u^+|^2\right)\notag\\
&+(m_{H_d}^2+|\lambda S|^2)\left(|H_d^0|^2+|H_d^+|^2\right)+\frac{g^2+g'^2}{8}\left(|H_u^0|^2+|H_u^+|^2-|H_d^0|^2-|H_d^-|^2\right)^2\notag\\
&+\frac{g^2}{2}|H_u^+H_d^{0*}+H_u^0H_d^{-*}|^2+m_S^2|S|^2+(\lambda A_{\lambda}(H_u^+H_d^--H_u^0H_d^0)S+\frac{1}{3}\kappa A_{\kappa}S^3+h.c).\label{HiggsPot}
\end{align}
Electroweak symmetry breaking occurs for appropriate values of the soft terms.
The Higgs fields are expanded around their vacuum expectation values,
\begin{align}
H_d&=\begin{pmatrix}v_d+\frac{h_d^0+ia_d^0}{\sqrt{2}}\\h_d^-\end{pmatrix},\\
H_u&=\begin{pmatrix}h_u^+\\v_u+\frac{h_u^0+ia_u^0}{\sqrt{2}}\end{pmatrix},\\
S&=s+\frac{h_s^0+ia_s^0}{\sqrt{2}}.
\label{eq:HiggsPot}
\end{align}
The vacuum expectation values, $v_u,v_d,s$ are chosen to be real and positive. As in the MSSM we define

\beqn
v^2=v_u^2+v_d^2,  \quad \tan\beta\equiv t_\beta=v_u/v_d, \quad (v_{u,d}=v s_\beta,v
c_\beta) 
\eeqn
such that the $W$ mass is
\beqn
M_W^2=g^2v^2/2 .
\eeqn

The non vanishing value of the vev of $S$  also gives a solution to the
so-called $\mu$-problem of the MSSM by generating this parameter dynamically:
\begin{equation}
\mu_{eff}=\lambda s.
\end{equation}
We define $\mu_{eff}=\mu$ in the following and will take it as an independent
parameter, comparison with the  MSSM will then be easier.  In addition to $\mu$,
we take $\lambda$ and $\kappa$ as independent parameters while $s$ is kept as a
shorthand notation for
$\mu/\lambda$ in the same way as we use $c_W$ as a short-hand notation for
$M_W/M_Z$. It is useful to introduce the combinations 
\beqn
\label{eq:mass_sing1}
\Lambda_v&=&\lambda v \quad {\rm and} \nonumber \\
m_\kappa &= &\kappa s=(\kappa/\l) \mu. 
\eeqn
 With these parameters, the MSSM limit is
obtained by taking $\kappa, \l (\Lambda_v)\to 0$, while keeping $\mu$  fixed such that the mass of the higgsinos is $|\mu|$. 
The reason we take $m_\kappa$ is that the mass of the singlino-like neutralino
is a substitute for $m_\k$. Indeed, in the MSSM limit, the singlino mass is,
see \cite{Belanger:2016tqb} 
\beqn
\label{eq:mass_sing2}
m_{\tilde{s}}= 2 m_\k.
\eeqn

At the minimum of the potential, the  part of the potential linear in any of the
CP-even Higgs field  has to  vanish, $\partial V_{H_{{\rm Higgs}}}/\partial
h_i^0=0$. It  can be written in terms of the tree-level tadpoles,

\begin{align}
\label{tad_tree_0}
\frac{\mathcal{T}_{h_d^0}}{2 v_d}=&-\mu \tb (A_\lambda+m_\kappa)+\Lambda_v^2  s_\beta^2+(m_{H_d}^2+\mu^2)+\frac{M_Z^2}{2} c_{2 \beta}, \nonumber \\
\frac{\mathcal{T}_{h_u^0}}{2 v_u}=&-\frac{\mu}{\tb} (A_\lambda+m_\kappa)+\Lambda_v^2 c_\beta^2+(m_{H_u}^2+\mu^2)-\frac{M_Z^2}{2} c_{2 \beta},\\
\frac{\mathcal{T}_{h_s^0}}{2 s}=&\quad m_\kappa(A_{\kappa}+2m_\kappa)+\Lambda_v^2 \big( 1 - s_{2\beta} \frac{A_\lambda+2 m_\kappa}{2 \mu}\big)+m_S^2. \nonumber
\end{align}

The  conditions on the  vanishing of the three tadpoles allow us to express the
soft mass terms $m_{H_d}^2,m_{H_u}^2, m_S^2$ in terms of the tadpoles.

The quadratic part of the Higgs potential, bilinear in the fields,  gives rise to the mass terms for the Higgs sector:
\begin{equation}
V_{mass}=\frac{1}{2} (h^0)^T \cdot M_S^2\cdot h^0
+\frac{1}{2} (a^0)^T \cdot M_P^2 \cdot a^0
+ h^- \cdot M_{\pm}^2 \cdot h^+ ,
\end{equation}
with 
\begin{align}
 (h^0)^T &= \begin{pmatrix}h_d^0&h_u^0&h_s^0\end{pmatrix} \\
 (a^0)^T &= \begin{pmatrix} a_d^0&a_u^0&a_s^0 \end{pmatrix} \\
 (h^\pm)^T &= \begin{pmatrix} h_d^\pm &h_u^\pm \end{pmatrix}
\end{align}
and where  $M_S^2$, $M_P^2$ and $M_{\pm}^2$ are respectively the mass matrices
for the CP-even,  the CP-odd, and the charged Higgs bosons.

\noindent
\underline{The charged Higgs}

The mass matrix for the charged Higgs reads
\begin{align}
\label{Mcharg}
M_{\pm}^2=\frac{1}{2}\begin{pmatrix}
\frac{\mathcal{T}_{h_d^0}}{v_d}&0\\
0&\frac{\mathcal{T}_{h_u^0}}{v_u}
\end{pmatrix} +
\left(\mu(A_{\lambda}+m_\kappa)+\frac{s_{2\beta}}{2}(M_W^2-\Lambda_v^2)\right)\begin{pmatrix}
t_\beta&1\\
1&1/t_{\beta}
\end{pmatrix},
\end{align}
With the tadpoles  set to zero (using the tree-level condition in Eq.~\ref{tad_tree_0}), we have ${\rm
Det}M_{\pm}^2=0$ which signals the presence of a massless charged Goldstone
boson. The mass of the physical charged Higgs boson is given by  the trace of
$M_{\pm}^2$,
\begin{equation}
\label{eq:mh+_tree}
M_{H^{\pm}}^2=\underbrace{\frac{2\mu}{s_{2\beta}}(A_\lambda+m_\kappa)}_{M_A^2=M_
{ A,{\rm MSSM}}^2}+\left(M_W^2-\Lambda_v^2\right).
\end{equation}

The MSSM limit is obtained by letting $\Lambda_v$ to $0$ in the above, while all
other parameters are fixed. What is denoted $M_A$ is  the equivalent of the
pseudoscalar mass in the MSSM limit. Note that if $m_\kappa (\kappa), \Lambda_v
(\lambda), \mu$ and $\tb$ have been extracted from the chargino/neutralino
sector, the measurement of the charged Higgs mass reconstructs $A_\lambda$.
As explained in \cite{Belanger:2016tqb}, for this to work efficiently $\tb$
should not be too large and in all cases should be well measured. The charged Higgs mass
could also serve for the measurement of $\tb$ if $A_\lambda$ is determined from
the other Higgs masses. 

The diagonalising matrix, $U(\beta)$, to obtain the Goldstone and physical
charged Higgs is defined as
\begin{align}
{{\cal H}}^\pm \equiv \begin{pmatrix}
G^{\pm}\\H^{\pm}
\end{pmatrix}=U_\beta h^\pm = U_\beta\begin{pmatrix}
h_d^{\pm}\\h_u^{\pm}
\end{pmatrix},
\end{align}
with 
\begin{equation}
\label{Ubeta}
U_\beta=\begin{pmatrix}
c_\beta&-s_\beta\\
s_\beta&c_\beta
\end{pmatrix}.
\end{equation}

\noindent
\underline{The Pseudoscalars}

The pseudoscalar mass matrix decomposes into the following elements
\begin{align}
\label{MPseudo}
\begin{array}{r c l}
M_{P_{11}}^2&=&\frac{\mathcal{T}_{h_d^0}}{2v_d}+\mu t_{\beta}(A_{\lambda}+m_\kappa),\\
M_{P_{22}}^2&=&\frac{\mathcal{T}_{h_u^0}}{2v_u}+ \frac{\mu}{t_{\beta}}(A_{\lambda}+m_\kappa),\\
\vspace{0.1cm}
M_{P_{33}}^2&=&\frac{\mathcal{T}_{h_s^0}}{2s}+ \Lambda_v^2 \frac{A_{\lambda}+4m_\kappa}{2 \mu}s_{2\beta}-3m_\kappa A_{\kappa},\\
\vspace{0.1cm}
M_{P_{12}}^2&=&M_{P_{21}}^2=\mu (A_{\lambda}+m_\kappa),\\
\vspace{0.1cm}
M_{P_{13}}^2&=&M_{P_{31}}^2=\Lambda_v (A_{\lambda}-2 m_\kappa) s_\beta,\\
M_{P_{23}}^2&=&M_{P_{32}}^2=\Lambda_v (A_{\lambda}-2m_\kappa) c_\beta,
\end{array}
\end{align}
As expected, upon setting the tadpole to zero,  ${\rm Det}M_P^2=0$. This reveals the neutral Goldstone boson.
With  the $2 \times 2$ submatrix, $m_{12}^2$ 
\begin{align}
m_{12}^2=s_\beta c_\beta M_{A,{\rm MSSM}}^2 \begin{pmatrix}
t_\beta&1\\
1&1/t_{\beta}
\end{pmatrix}, \quad {\rm Tr}(m_{12}^2)= M_{A,{\rm MSSM}}^2,
\end{align}
in the MSSM limit ($\Lambda_v \to 0$) we have 
\begin{align}
M_P^2 \to \begin{pmatrix}
m_{12}^2& 0 \\
0& -3 m_\kappa A_\kappa
\end{pmatrix}
\end{align}

The Goldstone boson can be isolated through the $3 \times 3$ extension of the
matrix $U_\beta$ encountered for the charged Higgs sector
\begin{equation}
U^{(3)}(\beta) =\begin{pmatrix}
c_\beta&-s_\beta&0\\s_\beta&c_\beta&0\\0&0&1
\end{pmatrix}.
\end{equation} 
In this new basis where the Goldstone boson is separated, the pseudoscalar mass matrix simplifies to  
\begin{equation}
U^{(3)}(\beta)  M_P^2 U^{(3)}(\beta) ^\dagger \; = \;
\begin{pmatrix}
\hspace*{-0.09cm}0 & \begin{matrix} 0 & 0 \end{matrix} \\
\begin{matrix} 0 \\ 0 \end{matrix}    & \scalebox{1.5}{$\widehat{M_{P}^2}$}
\end{pmatrix}
\end{equation}
where the ($2\times 2$) mixing matrix between the two pseudoscalar bosons is given by
\begin{equation}
\widehat{M_{P}^2} =\begin{pmatrix}
M_A^2&\Lambda_v (A_{\lambda}-2m_\kappa)\\
\Lambda_v (A_{\lambda}-2m_\kappa)&\Lambda_v^2 \frac{A_{\lambda}+4m_\kappa}{2 \mu}s_{2\beta}-3m_\kappa A_{\kappa}\end{pmatrix}.
\end{equation}

The MSSM limit is clearly exhibited. Diagonalisation of this matrix is then performed through a $2\times 2$ matrix 
$\hat{P_a}$ which we can parameterise as 

\beqn
\hat{P_a}=\begin{pmatrix}
c_p & -s_p\\
s_p & c_p
\end{pmatrix}
\eeqn
Putting everything together the pseudoscalar mass matrix is diagonalised
through the matrix $P_a$, 

\begin{align}
P_a= \underbrace{\begin{pmatrix}
\hspace*{-0.09cm}1 & \begin{matrix} 0 & 0 \end{matrix} \\
\begin{matrix} 0 \\ 0 \end{matrix}    & \scalebox{1.5}{$\hat{P_a}$}
\end{pmatrix} }_{\hspace*{0mm}\scalebox{1.2}{$\hat{P}_a^{(3)}$}}
U^{(3)}(\beta)=
\begin{pmatrix}
c_\beta & -s_\beta & 0 \\
  c_p s_\beta & c_p c_\beta & -s_p \\
  s_p s_\beta & s_p c_\beta & c_p
 \end{pmatrix}
\end{align}
such that

\begin{align}
{{\cal P}^0} \equiv
\begin{pmatrix}
G^0\\A_1^0\\A_2^0
\end{pmatrix}=P_a a^0 = P_a\begin{pmatrix}
a_d^0\\a_u^0\\a_s^0
\end{pmatrix},\quad 
\end{align}
It is important to remember that 

\beqn
\label{PaU3b}
(P_a)_{13}=(P_a^{-1})_{31}=0 \quad {\rm and \; that} \quad
(P_a)_{1i}=(U^{(3)}(\beta))_{1i} \quad {\rm for} \quad  i=1,2.
\eeqn

We will also set $({{\cal P}^0})_1\equiv A^0_0 \equiv G^0$ for the identification of the neutral Goldstone Boson.

\noindent
\underline{The CP-even scalars}

The elements of the scalar mass matrix read
\begin{align}
\label{Mheven}
\begin{array}{r c l}
M_{S_{11}}^2&=&\frac{\mathcal{T}_{h_d^0}}{2v_d}+M_Z^2 c_\beta^2+M_A^2 s_\beta^2,\\
M_{S_{22}}^2&=&\frac{\mathcal{T}_{h_u^0}}{2v_u}+M_Z^2 s_\beta^2+M_A^2 c_\beta^2,\\
\vspace{0.1cm}
M_{S_{33}}^2&=&\frac{\mathcal{T}_{h_s^0}}{2s}+\Lambda_v^2 A_{\lambda}\frac{c_\beta s_\beta}{\mu}+m_\kappa (A_{\kappa}+4m_\kappa),\\
\vspace{0.1cm}
M_{S_{12}}^2&=&M_{S_{21}}^2=\left(2\Lambda_v^2-M_Z^2-M_A^2\right)
s_\beta c_\beta,\\
\vspace{0.1cm}
M_{S_{13}}^2&=&M_{S_{31}}^2=\Lambda_v(2\mu c_\beta-(A_{\lambda}+2 m_\kappa )s_\beta),\\
M_{S_{23}}^2&=&M_{S_{32}}^2=\Lambda_v(2\mu s_\beta-(A_{\lambda}+2m_\kappa)c_\beta),
\end{array}
\end{align}
To get the physical eigenstates, we introduce  the orthogonal matrix $S_h$, such that 
\begin{align}
\begin{pmatrix}
h_1^0\\h_2^0\\h_3^0
\end{pmatrix}=S_h\begin{pmatrix}
h_d^0\\h_u^0\\h_s^0
\end{pmatrix} ,\quad 
\end{align}
We can make  the MSSM
limit  more apparent by writing the diagonalising matrix of the
CP-even Higgs mass matrix, $S_h=\hat{S_h} U^{(3)}(\beta)$. After rotation by
$U^{(3)}(\beta)$ an upper limit for
the non-singlet and CP-even lightest neutral Higgs mass is contained in a element of
the mass
matrix. The upper bound on this mass is 
$$M_{h_1^0}^2 < M_Z^2 \bigg( c_{2\beta}^2 +\frac{\Lambda_v^2}{M_Z^2} s_{2
\beta}^2 \bigg) \equiv M_Z^2 \bigg( 1+\bigg(\frac{\Lambda_v^2}{M_Z^2} -1 \bigg)
s_{2 \beta}^2 \bigg).$$
To have a tree-level mass which is higher than $M_Z$, $\Lambda_v$ needs to be
larger than $M_Z$ ($\lambda > \sqrt{g^2+g^{\prime 2}}/2$). Moreover the largest tree
level mass corresponds to moderate values of $\tb$. We need to
keep this in mind when we will discuss our benchmark points.

It is also useful to write 
\begin{equation}
M_{h_i^0}^2=\sum_{j,k=1}^3S_{h_{ij}}S_{h_{ik}}M_{S_{jk}}^2,\quad
M_{A_i^0}^2=\sum_{j,k=1}^3P_{a_{(i+1)j}}P_{a_{(i+1)k}}M_{P_{jk}}^2.
\label{HiggsMasses}
\end{equation} 

\noindent The properties of the physical states depend critically on the mixing
matrixes $S_h$ for the parity-even Higgses and on $P_a$ for the parity odd Higgses.
These mixing matrices which stem from the non-diagonal nature of the mass
matrices/bi-linear terms,  introduce a highly non linear dependence of the
couplings involving the Higgses on the underlying parameters of the theory 
whereas before mixing, in so to speak the current basis, the functional dependence of the Higgs couplings on the
underlying parameters is quite simple, linear or quadratic. This can be seen
from the Higgs potential in Eq.~\ref{eq:HiggsPot}.  For example,  before these rotation matrices are introduced, 
the coupling between 3 different CP-even neutral Higgses ($h_d^0 h_u^0 h_s^0$) is proportional to $\l
A_\l+2  \k \mu =\Lambda_v(A_\lambda+ 2 m_\kappa)/v$ and hence directly proportional to $\lambda$. After moving to the physical basis, the $h_1^0
h_2^0 h_3^0$ coupling is much more complicated since it involves the product of three
$S_h$. Therefore the dependence of this coupling on the underlying
parameters is more difficult to track. Since the triple Higgs
couplings will enter some of  the decays we will study, we write them below for the
CP-even Higgs sector,

\begin{eqnarray}
\label{hhhlamb}
\sqrt{2} v  g_{h_i^0 h_j^0 h_k^0} &=&
\frac{M_Z^2}{2}\left(\cosb(\Pi^S)^{1,1,1}_{i,j,k}+
\sinb(\Pi^S)^{2,2,2}_{i,j,k}\right)+
\left(\lv^2-\frac{M_Z^2}{2}\right)\left(\cosb(\Pi^S)^{1,2,
2}_{i,j,k} +\sinb(\Pi^S)^ { 2, 1,1 }_{ i,j,k}\right) \non \\
&+&\lv^2 \left( (\cosb  -  \frac{m_\kappa}{\mu}\sinb)(\Pi^S)^{1,3,3}_{i,j,k}
+(\sinb  -  \frac{m_\kappa}{\mu}\cosb)(\Pi^S)^{2,3,3}_{i,j,k} \right)  \\ 
&+& \Lambda_v \left( \mu \left((\Pi^S)^{3,1,1}_{i,j,k}+(\Pi^S)^{3,2,2}_{i,j,k}
\right)- (A_\lambda+ 2 m_\kappa )(\Pi^S)^{3,1,2}_{i,j,k} +
\frac{m_\kappa}{\mu} \frac{A_\k+6 m_\kappa}{3}(\Pi^S)^{3,3,3}_{i,j,k} \right) \non 
\end{eqnarray}
\noi
where the $\Pi^S$ represent the product of three $S_h$
\begin{equation}
(\Pi^S)^{a,b,c}_{i,j,k}=S_{h_{ia}}(S_{h_{jb}}S_{h_{kc}}+S_{h_{jc}}S_{h_{kb}})
+S_{h_{ib}}(S_{h_{ja}}S_{h_{kc}}+S_{h_{jc}}S_{h_{ka}})
+S_{h_{ic}}(S_{h_{ja}}S_{h_{kb}}+S_{h_{jb}}S_{h_{ka}})
\end{equation}\noi 
In the case where mixing is neglected in $S_h$ we have
\begin{eqnarray}
 (\Pi^S)^{a,b,c}_{i,j,k}&=& \d_{ia} (\d_{jb}
\d_{kc}+\d_{jc}\d_{kb})+\d_{ib}( \d_{ja} \d_{kc}+\d_{jc}
\d_{ka})+\d_{ic}( \d_{ja} \d_{kb}+ \d_{jb} \d_{ka}) \non \\
 (\Pi^S)^{a,b,c}_{i,i,i}&=& 6 \d_{ia} \d_{ib} \d_{ic} \non \\
 (\Pi^S)^{a,a,a}_{i,j,k}&=& 6 \d_{ia} \d_{ja} \d_{ka} \non \\
 (\Pi^S)^{a,b,b}_{i,j,k}&=& 2 \d_{ia} \d_{jb} \d_{kb} + 2 \d_{ib}
( \d_{ja} \d_{kb}+\d_{jb} \d_{ka}) \non 
\end{eqnarray}

It is important to realise that with our choice of the independent parameters
all triple Higgs couplings involving the singlet are proportional to $\lambda$
or $\lambda^2$. This should not be the case for the coupling between three
singlets which gets contributions from $S^3$ and $|S^2|^2$ terms. In
Eq.~\ref{hhhlamb} this is proportional to $\lambda m_\kappa/\mu=\kappa$. The
fact that this coupling exhibits a $\lambda$ dependence is due to our choice of
inputs $m_\kappa$ and $\mu$ which are more directly related to the mass of the
singlino and the higgsino, see Eqs.\ref{eq:mass_sing1}.

\subsection{Counting parameters and fields}
\label{subsec:counting_para}
Let us take stock and summarise the situation as regards the number of (physical) parameters and fields in the Higgs sector of the NMSSM. 
The physical scalar fields consist of 3 neutral CP-even Higgs bosons, $h_1^0,h_2^0,h_3^0$, 2 CP-odd Higgs bosons,
$A_1^0,A_2^0$ and a charged Higgs boson, $H^\pm$.  The NMSSM contains of course
the SM gauge fields (and fermions). In particular the SM gauge parameters 
\beqn
\label{para-ggv}
g, g^\prime \quad  {\rm and} \quad v=v_u^2+v_ d^2
\eeqn
are traded for the following physical input parameters 
\beqn
\label{para-mwmze}
e,M_W,M_Z
\eeqn
For these parameters we will apply the usual On-Shell (OS) renormalisation scheme. In particular $e$ will be defined in the Thomson limit. The Thomson limit, $q^2 \to 0$, may not be the most appropriate scale for the NMSSM processes whose loop corrections we will study, however one can easily quantify the effect of using a running $\alpha_{{\rm e.m.}}$ at the scale of the process. Besides these  standard model parameters, the NMSSM introduces an additional set of  9 parameters  from the Higgs sector alone. 
From the Higgs potential, equation (\ref{HiggsPot}), it is clear  that the Higgs sector of the NMSSM depends on the parameters:

\begin{equation}
\label{para_count_higgs_1}
\underbrace{t_\beta, \lambda,\kappa,\mu}_{{\rm in} \; \tilde{\chi} \; {\rm sector \; also}}, A_{\lambda}, A_\kappa,
m_{H_d},m_{H_u},m_{S}.
\end{equation}
where the first four parameters are also involved in the characterisation of the
neutralino/chargino sector which we studied at length in the sister paper~\cite{Belanger:2016tqb}.
Alternatively the last three soft Higgs masses can be traded for the tadpoles of
 the neutral Higgs which need to be constrained to zero to impose that the
potential is at its minimum. The latter are therefore considered as  physical
observables. 
\begin{equation}
\label{para_count_higgs_2}
\underbrace{t_\beta, \lambda,\kappa (m_\k),\mu}_{{\rm in} \; \tilde{\chi} \; {\rm sector \; also}}, A_{\lambda}, A_\kappa,
\left({{\cal T}}_{H_d},{{\cal T}}_{H_u},{{\cal T}}_{S}\right).
\end{equation}
The first six parameters above are not 
unambiguously defined in a simple mapping to an observable. We will discuss at
length the choice and definitions of the input parameters that will construct
the set of these six parameters. This issue is directly related to the
renormalisation scheme. 

The parameters and fields of the neutralino/chargino sector were described in ~\cite{Belanger:2016tqb}.   The parameters of this sector are the $U(1)$ and $SU(2)$ gaugino soft masses, $M_1,M_2$ in addition  to $t_\beta, \lambda,\kappa,\mu$. 
\\

\section{Renormalisation of the Higgs sector}
\label{sec:ren_higgs}
\subsection{A word about the gauge-fixing}
In this work we have restricted ourselves to the simplest gauge-fixing, namely
we take a linear gauge fixing constraint with a 'tHooft-Feynman parameter set to
$1$. Only the SM fields (including the necessary) Goldstone bosons appear,
namely 
\beqn
\label{eq:GF}
{{\cal L}}_{\rm GF}= - |\partial^\mu W_{\mu}^++i M_W G^+|^2
-\frac{1}{2}|\partial^\mu Z_{\mu }^0+i \frac{M_Z}{2} G^0|^2- |\partial^\mu
A_{\mu}|^2.
\eeqn
It is important to stress that all fields and parameters in Eq.~\ref{eq:GF} are understood to be {\em renormalised}. 

\subsection{Parameters, fields and self-energies}
What is also considered as renormalised are all the rotation matrices $U(\beta),
P_a, S_h$. This is an approach we have consistently applied in all our work on
the renormalisation of supersymmetric models starting from the MSSM and imposed
in all the sectors of the models where mixing between fields occurs, not only in
the Higgs sector, but also in the sfermion, neutralino, chargino sectors
\cite{Baro:2008bg,Baro:2009gn,Belanger:2016tqb}. While one-loop corrections do
reintroduce mixing, the use of wave function renormalisation constants, with
judicious choices of conditions imposed at the physical masses of the particles,
will ensure that even at one-loop transitions between particles with the same
quantum numbers will vanish when these particles are on their mass shell. \\

\noi With the exception of ${{\cal L}}_{\rm GF}, U(\beta), P_a$ and $S_h$ all fields
and parameters encountered so far are bare quantities. All bare quantities
($X_0$) are then decomposed into renormalised ($X$) and counterterms ($\delta
X$) quantities. First,  the SM parameters are shifted such that 
\beqn
g \to g +\delta g \quad g^\prime \to g^\prime +\delta g^\prime \quad v \to v+\delta v, 
\eeqn
which tantamount to 
\beqn 
\label{shifts_para_sm}
 e \to e +\delta e, \quad M_W \to M_W +\delta M_W, \quad M_Z \to M_Z +\delta
M_Z.
\eeqn

The same  procedure applies to the NMSSM parameters in
Eq.~\ref{para_count_higgs_1} or equivalently 
Eq.~\ref{para_count_higgs_2} with 
\beqn
\label{shifts_para_nmssm}
t_\beta, \lambda,m_\kappa,\mu, A_{\lambda}, A_\kappa \non & \to & t_\beta
+\delta
t_\beta, \lambda +\delta \lambda,m_\kappa+\delta m_\kappa ,\mu+\delta \mu, 
A_{\lambda}+\delta A_{\lambda}, A_\kappa+\delta A_\kappa, \non \\
{{\cal T}}_{H_d},{{\cal T}}_{H_u},{{\cal T}}_{S} & \to &
{{\cal T}}_{H_d}+\delta {{\cal T}}_{H_d},{{\cal T}}_{H_u}+\delta{{\cal T}}_{H_u},{{\cal T}}_{S}+{{\cal T}}_{S}.
\eeqn

For the fields  all shifts are directly encoded in the wave function renormalisation constants, so that mass eigenstates are expressed in terms of the gauge eigenstates in the same manner at one-loop and at tree-level. 
For the gauge fields we perform $W_\mu \to Z_W^{1/2} W_\mu$ while the system
$(A_\mu, Z_\mu^0)$ involves a matrix with 4 entries $\delta Z_{\gamma \gamma},
\delta Z_{\gamma Z},\delta Z_{Z \gamma},\delta Z_{ZZ}$, see
\cite{Belanger:2003sd}. 
For the NMSSM Higgs sector, this entails that the three wave-function renormalisation matrices $Z_S$, $Z_P$ and $Z_C$, are  introduced such that,
\begin{equation}
\begin{pmatrix}
h_1^0\\h_2^0\\h_3^0
\end{pmatrix}_0
=Z_S^{1/2}
\begin{pmatrix}
h_1^0\\h_2^0\\h_3^0
\end{pmatrix},\quad \begin{pmatrix}
G^0\\A_1^0\\A_2^0
\end{pmatrix}_0
=Z_P^{1/2}
\begin{pmatrix}
G^0\\A_1^0\\A_2^0
\end{pmatrix},\quad
\begin{pmatrix}
G^\pm\\H^\pm
\end{pmatrix}_0=Z_C^{1/2}\begin{pmatrix}
G^\pm\\H^\pm
\end{pmatrix}
\end{equation}
where the index 0 is attached to the bare fields while  the renormalised fields do not have an index. The elements of the wave-function renormalisation matrices can be written as, 
\begin{equation}
Z_C^{1/2}=\begin{pmatrix}
1+\frac{1}{2}\delta Z_{G^\pm}&\frac{1}{2}\delta Z_{G^\pm H^\pm}\\
\frac{1}{2}\delta Z_{H^\pm G^\pm}&1+\frac{1}{2}\delta Z_{H^\pm}
\end{pmatrix},
\end{equation}

\begin{equation}
Z_P^{1/2}=\begin{pmatrix}
1+\frac{1}{2}\delta Z_{G^0}&\frac{1}{2}\delta Z_{G^0 A_1^0}&\frac{1}{2}\delta Z_{G^0 A_2^0}\\
\frac{1}{2}\delta Z_{A_1^0 G^0}&1+\frac{1}{2}\delta Z_{A_1^0}&\frac{1}{2}\delta Z_{A_1^0 A_2^0}\\
\frac{1}{2}\delta Z_{A_2 G^0}&\frac{1}{2}\delta Z_{A_2^0 A_1^0}&1+\frac{1}{2}\delta Z_{A_2^0}
\end{pmatrix},
\end{equation}

\begin{equation}
\label{eq:Z_S}
Z_S^{1/2}=\begin{pmatrix}
1+\frac{1}{2}\delta Z_{h_1^0}&\frac{1}{2}\delta Z_{h_1^0h_2^0}&\frac{1}{2}\delta Z_{h_1^0h_3^0}\\
\frac{1}{2}\delta Z_{h_2^0h_1^0}&1+\frac{1}{2}\delta Z_{h_2^0}&\frac{1}{2}\delta Z_{h_2^0h_3^0}\\
\frac{1}{2}\delta Z_{h_3^0h_1^0}&\frac{1}{2}\delta Z_{h_3^0h_2^0}&1+\frac{1}{2}\delta Z_{h_3^0}
\end{pmatrix}.
\end{equation}

\subsection{One-point functions and tadpoles}
Once shifts on all parameters of the models 
including the tadpole terms (Eq.~\ref{shifts_para_nmssm}) and wave function renormalisation of the fields have
been performed, we  concentrate on the terms which are linear in the scalar
fields and combine them with the one-loop contribution to these tapdoles. The
tree-level condition on the tadpole is now elevated to the one-loop level so
that minimisation of the potential is realised. 
 At one-loop, the linear part of the potential can be written:
\begin{equation}
V_{lin}^{(1)}=(-\mathcal{T}_{h_d^0}^{loop}+\delta\mathcal{T}_{h_d^0})\frac{h_d^0}{\sqrt{2}}+(-\mathcal{T}_{h_u^0}^{loop}+\delta\mathcal{T}_{h_u^0})\frac{h_u^0}{\sqrt{2}}+(-\mathcal{T}_{h_s^0}^{loop}+\delta\mathcal{T}_{h_s^0})\frac{h_s^0}{\sqrt{2}},
\end{equation}
where the first terms in the parenthesis are the pure loop contributions and the
second ones the counterterms. We observe that because of the condition on the
tree-level tadpoles, wave function renormalisation of the Higgses does not
enter. Our first renormalisation condition is that these linear terms cancel.
The loop contributions for the gauge eigenstates tadpoles are obtained from the
mass eigenstates tadpoles with the use of the diagonalisation matrix $S_h$,
\begin{equation}
\begin{pmatrix}
\mathcal{T}_{h_d^0}^{loop}\\\mathcal{T}_{h_u^0}^{loop}\\\mathcal{T}_{h_s^0}^{loop}
\end{pmatrix}=S_h^{-1}\begin{pmatrix}
\mathcal{T}_{h_1^0}^{loop}\\\mathcal{T}_{h_2^0}^{loop}\\\mathcal{T}_{h_3^0}^{loop}
\end{pmatrix}.
\end{equation}
The minimum condition then gives simply,
\begin{equation}
\label{eq:tadpole_ct}
\delta\mathcal{T}_{h_i^0}=\mathcal{T}_{h_i^0}^{loop},\quad i=d,u,s.
\end{equation}

\section{Bilinears and two-point functions self-energies}
\label{sec:self-energies}
\subsection{Mass counterterms for the Higgs sector}
We now turn to the bilinear terms in the Higgs fields and perform shifts in the
parameters according to Eq.~\ref{shifts_para_sm} and
Eq.~\ref{shifts_para_nmssm}. These shifts are performed on each of the
underlying parameters (including the tadpoles) of the mass matrices  $M_\pm^2,
M_P^2, M_S^2$ in Eqs.~\ref{Mcharg},\ref{MPseudo},\ref{Mheven}. Since our
approach is to use the same tree-level diagonalising mass matrices, namely
$U_\beta,P_a,S_h$, to convert to the physical fields, after the shifts the
{\em ``physical"} fields will now mix. Therefore apart from the induced
diagonal counterterms $\delta M_{H^+}^2, \delta M_{A^0_{1,2}}^2, \delta
M_{h^0_{1,2,3}}^2$, spurious counterterms to the Goldstones $\delta
M_{G^0,G^\pm}^2$ are generated as well as non-diagonal mass mixings such as
$\delta M_{h_1^0h_2^0}^2$ and transitions such as $\delta M_{G^\pm H^\pm}^2$.
Therefore, we need to enforce appropriate conditions to the one-loop wave
function renormalisation matrices such that a correct on-shell definition and
normalisation of the external particles is ensured. This is obtained by imposing
that the residue of each (diagonal) propagator is equal to 1 (as is done in any
theory without mixing). Strictly speaking, if we were only interested in having
finite $S$-matrix elements and not finite Green's functions, wave function
renormalisation would not be a must. Still, the residues of the propagators of
the external particles must be set to 1 to correctly normalise the S-matrix, this can be  achieved by introducing finite wave function correction normalisation
factors to prevent non-vanishing transitions on the external legs. \\

\noi Since the contributions of tadpoles is very important, let us summarise how the
shifts in the underlying parameters affect the mass matrices. Generically the
mass matrix in the three sectors ($M^2=M_\pm^2, M_P^2, M_S^2$) can, in
the bases $d,u,s$, be decomposed into a diagonal tadpole matrix $T_M$ and a
non-tadpole matrix which we denote $M_M^2$, such that $M^2 = T_M + M_M^2$. At
tree-level all $T_M$ are set to zero and the diagonalising matrix $U$
($U=U_\beta, P_a, S_h$) is such that $M_D^2=U  M_M^2 U^{-1}$ is diagonal with
eigenvalues being the tree-level physical masses. In our notation, for the
charged and pseudoscalar sectors the Goldstones are the $(11)$ entries, hence
${(M_D)}_{11}=0$. The shifts entail the counterterm mass matrix
\beqn
\delta M^2=\delta T_M +\delta M_M^2.
\eeqn

In our approach, in all sectors of the NMSSM, to move to the  ({\em
``physical"}) basis $A^0_i,G^0_i,\cdots$ we use the \underline{same matrix} as
the one at tree-level, namely (the sub-index $Ph.$ below generically denotes the
mass matrix in the ``physical`` basis)

\beqn
\label{ct_mass_matrix_code}
\delta M^2_{Ph.}&=& U\;\delta T_M \;U^{-1} + U\;\delta M_M^2\;U^{-1}.
\eeqn
These are the mass counterterms which will be used to define the self-energies.
In the code these counterterms are generated according to  
Eq.~\ref{ct_mass_matrix_code}. However, for the sake of the discussion, it is
more enlightening to express the above counterterms through the mass
eigenvalues, $M_D^2=U  M_M^2 U^{-1}$ (note that there is no tadpole here). It is
important to realise that shifts in $M_D$ now
should be understood as shifts in {\em all parameters} that define $M_D$ and in
particular those entering $U$. We can then write, using the unitarity of $U$,
$\delta (U) U^{-1} +U \delta (U^{-1})=0$

\beqn
\label{ct_mass_matrix_code2}
\delta M^2_{Ph.}&=& U\;\delta T_M \;U^{-1} +\delta M_D^2 + \left( U (\delta
U^{-1}) M_D^2 + M_D^2 (\delta U) U^{-1}) \right) \non \\
(\delta M^2_{Ph.})_{ij}&=&(U\;\delta T_M \;U^{-1})_{ij} +(\delta M_D)_{i}^2
\delta_{ij} +\left(\delta (U)  U^{-1}\right)_{ij}
\left((M_D)_i^2-(M_D)_j^2\right)
\eeqn

Eq.~\ref{ct_mass_matrix_code2} shows that for the diagonal terms, the terms in
$\delta U$ vanish, in particular the counterterms to the Goldstones for both  the
charged and the pseudoscalar, $\delta M_{G}^2$ (for $i=j=1$), is a pure tadpole
term as it should.  From the properties of the $P_a$ matrix of the
pseudoscalars we even find that $\delta M_{G^{\pm}}^2=\delta M_{G^{0}}^2$. The
mixing between a Goldstone and a non Goldstone field is proportional to the
(tree-level) mass of the associated non Goldstone physical field. Because
$\delta M^2_{Ph.}$ is no longer diagonal, Goldstone-Higgs mixing mass
counter-terms are generated, 
\beqn
\delta M_{H^\pm G^\mp}^2&=&\delta M_{G^\mp H^\pm }^2=\frac{s_{2\beta}}{2}\left(
\delta T + M_{H^\pm}^2 \frac{\delta t_\beta}{t_\beta}\right) \\
\d M_{A_i^0 G^0}^2 &=& \d M_{G^0 A_i^0}^2 = \frac{\sbt}{2}\left(\delta T
 + M_{A_i^0}^2 \frac{\delta t_\beta}{t_\beta}\right) 
\eeqn
where
\begin{equation}
 \d T = \frac{1}{v \sbt} \sum_{i=1}^3 \left(\sinb S_{h,i1}- \cosb
S_{h,i2}\right)\d {\cal T}_{h_i^0}
\end{equation}\noi 
and the need for the wave function renormalisation becomes evident in order to counterbalance the appearance of these transitions especially when the particles are on their mass shell. 

\subsection{Two-point functions from the Higgs potential}
Implementing the wave function renormalisation directly in
the {\em physical} basis, we can write the renormalised self-energies, with the
non ``hatted" expression as the result of the 1-loop unrenormalised
self-energy, while the $\delta Z$'s are the result of the wave function
renormalisation.  The mass shifts correspond exactly to the elements of
$\delta M^2_{Ph.}$ (which include tadpoles). For the CP-even scalars we obtain
$(i,j=1,2,3)$,
\begin{equation}
\label{CPevenSelf}
\begin{array}{r c l}
 \hat{\Sigma}_{h_i^0h_j^0}(p^2)&=&\Sigma_{h_i^0h_j^0}(p^2)+\delta
M_{h_i^0h_j^0}^2-\frac{1}{2}(p^2-M_{h_i^0}^2)\delta
Z_{h_i^0h_j^0}-\frac{1}{2}(p^2-M_{h_j^0}^2)\delta Z_{h_j^0h_i^0} 
\end{array}
\end{equation}\noi 
while for the CP-odd scalars we get $(i,j=1,2)$,
\begin{equation}
\label{CPoddSelf}
\begin{array}{r c l}
 \hat{\Sigma}_{G^0G^0}(p^2)&=&\Sigma_{G^0G^0}(p^2)+\delta M_{G^0}^2-p^2\delta
Z_{G^0}\\
\hat{\Sigma}_{A_i^0G^0}(p^2)&=&\Sigma_{A_i^0G^0}(p^2)+\delta
M_{A_i^0G^0}^2-\frac{1}{2}p^2\delta
Z_{G^0A_i^0}-\frac{1}{2}(p^2-M_{A_i^0}^2)\delta Z_{A_i^0G^0}\\
\hat{\Sigma}_{A_i^0A_j^0}(p^2)&=&\Sigma_{A_i^0A_j^0}(p^2)+\delta
M_{A_i^0A_j^0}^2-\frac{1}{2}(p^2-M_{A_i^0}^2)\delta
Z_{A_i^0A_j^0}-\frac{1}{2}(p^2-M_{A_j^0}^2)\delta Z_{A_j^0A_i^0}
\end{array}
\end{equation}
and the charged scalars,
\begin{equation}
\label{ChargedSelf}
 \begin{array}{r c l}
\hat{\Sigma}_{G^\pm G^\pm}(p^2)&=&\Sigma_{G^\pm G^\pm}(p^2)+\delta M_{G^\pm
}^2-p^2\delta Z_{G^\pm}\\
\hat{\Sigma}_{G^\pm H^\pm}(p^2)&=&\Sigma_{G^\pm H^\pm}(p^2)+\delta M_{G^\pm
H^\pm}^2-\frac{1}{2}p^2\delta Z_{G^\pm H^\pm}-\frac{1}{2}(p^2-M_{H^\pm}^2)\delta
Z_{H^\pm G^\pm}\\
\hat{\Sigma}_{H^\pm H^\pm}(p^2)&=&\Sigma_{H^\pm H^\pm}(p^2)+\delta M_{H^\pm
}^2-(p^2-M_{H^\pm }^2)\delta Z_{H^\pm}
\end{array}
\end{equation}

The aim now is to determine all the counterterms entering these expressions. Note that the Goldstones will mix and we have singled out their appearance. Recall that
the Goldstone bosons are not physical, they cannot appear in initial or final states of a physical process. Thus, we do not need to renormalise their wave function and we can set $\delta Z_{G^0}=\delta Z_{G^\pm}=0$. Moreover, $\delta Z_{A_i^0 G^0}$ and $\delta Z_{H^\pm G^\pm}$ can also be set to 0 since they also correspond to transitions where a Goldstone boson is on an external leg.

\subsection{$H^\pm W^\pm$ and $A^0_{1,2} Z^0$ transitions}
At tree-level the gauge fixing eliminates mixing between the gauge bosons and
their corresponding Goldstone bosons therefore compensating for such a mixing
that emerges from the gauge sector, ${\cal L}^{GV}$ (the gauge covariant kinetic
term of the Higgs fields). We follow an approach where the gauge-fixing is
unrenormalised. As a result of shifting (both fields and parameters) in ${\cal
L}^{GV}$, the massive gauge bosons and the pseudoscalars (as well as the
Goldstones) will mix. Gauge invariance relates the gauge-pseudoscalar
transitions and the corresponding Goldstones-pseudoscalar transition which we
studied previously. Therefore we need to consider these transitions. 

To get the remaining counterterms involving Goldstones, $\delta Z_{G^0A_i^0}$
and $\delta Z_{G^\pm H^\pm}$, one also has to deal with new transitions between
gauge bosons and CP-odd or charged Higgses.
The expansion of the covariant derivative in the kinetic part of the scalar
Lagrangian gives the following interaction terms,
\begin{align}
{\cal L}^{GV}= M_Z\left(\cosb\partial^\mu a_d^0- \sinb\partial^\mu
a_u^0\right)Z_\mu^0 - M_W\left( i \left(\cosb \partial^\mu
h_d^- -\sinb \partial^\mu h_u^-\right)W_\mu^++h.c\right)
\label{LagrGV}
\end{align}

At tree-level the combination of the scalar fields makes up the Goldstone bosons, the first component of the corresponding Higgs fields, namely 
\beqn
\cosb a_d^0- \sinb a_u^0 & \equiv &(P_a a^0)_1=({\cal {P}}^0)_1=G^{0} \nonumber
\\
\cosb h_d^- -\sinb  h_u^- & \equiv & (U(\beta) h^-)_1=({\cal {H}}^-)_1=G^-
\eeqn

Take for example, the case of the pseudoscalar/neutral Goldstones. Before applying the wave function renormalisation, the shifts amount to 
 
 \beqn
 (P_a a^0)_1 \to ((\delta P_a) a^0 )_1= \big((\delta P_a) P_a^{-1} {\cal {P}}^0) \big)_1=\big((\delta P_a) P_a^{-1} \big)_{1i} ({\cal {P}}^0)_i 
 \eeqn
This shift alone, prior to wave  function renormalisation will introduce $A_i^0
Z^0$ transitions, but not $G^0 Z^0$. This is easy to see. With
$P_a=\hat{P}_a^{(3)} U(\beta)^{(3)}$ and using the fact that
$\left(\hat{P}_a^{(3)}\right)_{1i}=\delta_{1i}$ allows to write

\beqn
\sum_{i=1}^{3}\big((\delta P_a) P_a^{-1} \big)_{1i} ({\cal
{P}}^0)_i&=&-\sum_{i=1}^{2} A_i^0 \frac{s_{2\beta}}{2} 
\Big(c_\beta P_{a,(i+1)2}+s_\beta P_{a,(i+1)1}\Big)\frac{\delta
t_\beta}{t_\beta} \nonumber \\
&=&-\sum_{i=1}^{2} A_i^0 \frac{s_{2\beta}}{2} 
\hat{P}_{a,i1} \frac{\delta
t_\beta}{t_\beta}\eeqn

Including the wave function renormalisation, we obtain 

 \beqn
 \delta{{\cal L}}_{{\rm neutral}}^{GV}=\frac{M_Z}{2}&\Big\{&\Big(\d Z_{ZZ}+\d
Z_{G^0}+\frac{\d M_Z^2}{M_Z^2} \Big)\partial^\mu G^0 Z_\mu^0+
\d Z_{Z\gamma} \partial^\mu G^0 A_\mu \nonumber \\
&+& \sum_{i=1}^{2}\Big(\d Z_{G^0 A_i^0}-\sbt
\hat{P}_{a,i1} \frac{\delta
t_\beta}{t_\beta}\Big) \partial^\mu A_i^0 Z_\mu^0 \Big\}
\eeqn
\noi 
The first  line of the equation above is the usual SM term. 
However, there remains non-vanishing transitions between pseudoscalars
and the
$Z^0$ boson, leading to the following self-energies :
\begin{align}
\label{dzga0}
\hat{\Sigma}_{A_i^0 Z^0}(p^2)=&\Sigma_{A_i^0 Z^0}(p^2)+\frac{M_Z}{2}\Big(\delta
Z_{G^0
A_i^0}-s_{2\beta} \hat{P}_{a,i1}\frac{\delta
t_\beta}{t_\beta}\Big).
\end{align}
Following the same steps,  the transition between
the charged Higgs and the $W^\pm$ boson is given by
\begin{equation}
\label{dzghp}
\hat{\Sigma}_{H^\pm W^\pm}(p^2)=\Sigma_{H^\pm
W^\pm}(p^2)+\frac{M_W}{2}\Big(\delta Z_{G^\pm H^\pm}-s_{2\beta}\frac{\delta
t_\beta}{t_\beta}\Big).
\end{equation}

In the linear gauge we have used, there is a simple Ward identity which was
derived in \cite{Baro:2008bg}. It can be readily extended to the NMSSM using the
unitarity properties of the matrix $\hat{P}_a$. The starting point is  the BRST variation on the correlator between the Z-boson ghost and the pseudoscalar  $<0|\bar c_Z(x) A^0_i(y)|0>$. This induces an   identity which sets the
following strong constraints

\beqn
p^2\hat{\Sigma}_{H^\pm W^\pm}(p^2)+M_{W^\pm}\hat{\Sigma}_{H^\pm G^\pm}(p^2)=-\frac{M_{W^\pm}}{2} \left(p^2-M_{H^\pm}^2\right)
\Big(\delta Z_{ H^\pm G^\pm}+s_{2\beta}\frac{\delta
t_\beta}{t_\beta}- {\cal F}^\pm(p^2)\Big) \nonumber \\
p^2\hat{\Sigma}_{A^0_i Z^0}(p^2)+M_{Z^0}\hat{\Sigma}_{ A^0_i G^0}(p^2)=-\frac{M_{Z^0}}{2} \left(p^2-M_{A_i^0}^2\right)
\Big(\delta Z_{ A^0_i G^0}+s_{2\beta}   \hat{P}_{a,i1}    \frac{\delta
t_\beta}{t_\beta} - {\cal F}^0(p^2)\Big) \nonumber
\eeqn
where
\begin{align}
 {\cal F}^\pm (p^2) &= \frac{\alpha}{8 \pi s_W^2} \sum_{i=1}^3
\left(\cbt S_{h,i1}S_{h,i2} + \frac{\sbt}{2}(S_{h,i2}^2-S_{h,i1}^2)\right)
B_0(p^2,M_W^2,M_{h_i^0}^2) \\
 {\cal F}^0(p^2) &= \frac{\alpha }{2\pi s_{2W}^2}\hat{P}_{a,i1} \sum_{i=1}^3
\left(\cbt S_{h,i1}S_{h,i2} + \frac{\sbt}{2}(S_{h,i2}^2-S_{h,i1}^2)\right)
B_0(p^2,M_Z^2,M_{h_i^0}^2)
\end{align}\noi 
with $B_0(p^2,M_V^2,M_{h_i^0}^2)$ the scalar two point function
\cite{Passarino:1978jh} and $V = W,Z$.

The importance of these identities is that the $H^\pm G^\pm$ and $H^\pm W^\pm$
transitions (and their neutral counterparts) are not independent. In particular
we will be interested in setting an on-shell renormalisation scheme whereby, on
the mass-shell, a transition between the charged Higgs and the charged
Goldstone, and  any of the neutral pseudoscalars and the neutral  Goldstone
boson, does not occur at one-loop. In doing so, the identification
that is made for these states at tree-level is maintained. The previous
identities show that at $p^2=M_{H^\pm}^2$ and $p^2=M_{A_i^0}^2$, transitions
between these physical scalars and the corresponding gauge bosons do not occur
either. It means that one can simultaneously set

\beqn
\label{set_cdts_hw}
\hat{\Sigma}_{H^\pm W^\pm}(M_{H^\pm}^2)=\hat{\Sigma}_{H^\pm
G^\pm}(M_{H^\pm}^2)=0, \nonumber \\
\hat{\Sigma}_{A_i^0 Z^0}(M_{A_i^0}^2)=\hat{\Sigma}_{A_i^0
G^0}(M_{A_i^0}^2)=0
\eeqn

From Eqs~(\ref{dzga0}~,~\ref{dzghp}~,~\ref{set_cdts_hw}) we derive 
\beqn
\delta Z_{G^\pm H^\pm }&=& s_{2\beta}\frac{\delta
t_\beta}{t_\beta}- \frac{2}{M_{W^\pm}}\Sigma_{H^\pm
W^\pm}(M_{H^\pm}^2) \nonumber \\
\delta Z_{G^0A_i^0}&=&s_{2\beta}\hat{P}_{a,i1}\frac{\delta
t_\beta}{t_\beta}-\frac{2}{M_Z}\Sigma_{A_i^0 Z^0}(M_{A_i^0}^2)
\eeqn

The Higgs masses which appear as arguments of the two-point function in the
equations above are taken as the tree-level masses in order to be consistent with a fully
one-loop treatment.  These equations allow to fix all wave function
renormalisation constants pertaining to the Goldstone bosons, this will then 
leave us to deal with the system of the physical Higgses: the charged Higgs, the
pair of  pseudoscalars $A_i^0$ and the three CP-even neutral Higgses $h_i^0$.
Likewise with the gauge bosons whose renormalisation goes now exactly the same
way as the renormalisation of the gauge sector within the SM. For the latter we
follow the on-shell scheme adopted in \cite{Belanger:2003sd}. Note that
although this procedure permits to decouple the Goldstones from the physical
fields, to fully determine the values of $\delta Z_{G^\pm H^\pm}$ and $\delta
Z_{G^0A_i^0}$ we still need to define a renormalisation condition on $\delta
t_\beta$. This will need an input from the physical Higgses to which we now
turn.

\subsection{Renormalisation conditions from the Higgs self-energies}
With the Goldstone bosons now set aside, the renormalised self energies of the
Higgses in Eqs.~(\ref{CPevenSelf}, \ref{CPoddSelf} and \ref{ChargedSelf}) take
the same form, allowing for one-loop transitions between them. Again, we require
that the mixing between any two particles of the same CP parity must vanish at
the mass of any physical particle (on-shell condition). The mass is defined as
the pole mass of the real part of the renormalised inversed propagator. In case
of mixing this requires solving a $3\times3$ (for the CP-even) and $2\times 2$ (for the CP-odd) Higgses
system of an inverse propagator. 
At one-loop, these equations are linearised (see \cite{Baro:2008bg}). In this
case, starting with a tree-level Higgs mass, $M_{i,{\rm tree}}$ ($i$ generically
denotes $h^0_{1,2,3},A^0_{1,2},H^\pm$), the corrected one-loop mass is the
solution of the equation 

\beqn
p^2-M_{i,{\rm tree}}^2-{\rm Re}\hat{\Sigma}_{ii}(p^2)=0 \quad p^2=M_{i,{\rm 1loop}}^2,
\eeqn
which, in the one-loop approximation, reads
\beqn
M_{i,{\rm 1loop}}^2=M_{i,{\rm tree}}^2+{\rm Re}\hat{\Sigma}_{ii}(M_{i,{\rm tree}}^2)=
M_{i,{\rm tree}}^2+\delta M^2_i+{\rm Re}{\Sigma}_{ii}(M_{i,{\rm tree}}^2)
\eeqn

The above equation can be used in two ways. If all counterterms entering in
$\delta M^2_i$ have been fixed (and hence are known), the above equation
calculates the  {\em finite} correction to the tree-level mass. The ultraviolet
finiteness of the corrected mass  is a very powerful check on the implementation
of the one-loop set-up. We may also use one or some of the masses of the Higgses
as input parameters in order to solve for  one or some of the counterterms to the 
underlying parameters that enter in $\delta M_i^2$. In this case 
\beqn
M_{i,{\rm 1loop}}^2=M_{i,{\rm tree}}^2 \equiv M_{i,{\rm input}}^2 \longrightarrow \delta M_{i}^2=-{\rm Re}\Sigma_{ii}(M_{i,{\rm tree}}^2)
\eeqn
For instance taking the charged Higgs mass $M_{H^\pm}$ as an input parameter
gives
\beqn
\label{eq:input_mhp}
\delta M_H^\pm=-\Sigma_{H^\pm H^\pm}(M_{H^\pm}^2)
\eeqn
We also impose that the residue of the propagators for an on-shell physical field be equal to one such that
\beqn
{\rm Re}\hat{\Sigma}_{ii}^{\prime}(M_{i,{\rm tree}}^2)=1 \quad {\rm with} \quad
\frac{\partial \hat{\Sigma}_{ii}^{\prime}(p^2)}{\partial
p^2}=\hat{\Sigma}_{ii}^{\prime}(p^2)
\eeqn
which then fixes the diagonal entries of the wave function renormalisation constants such that 
\beqn
\delta Z_i={\rm Re} \Sigma_{ii}^{\prime}(M_{i,{\rm tree}}^2) \quad \left( {\rm for \; example} \; \delta Z_{h_i^0}={\rm Re}\Sigma_{h_i^0h_i^0}^{\prime}(M_{h_i^0}^2) \right) 
\eeqn

We also impose that no mixing occurs between two same-parity field when any of
them is on-shell. This condition translates into 
\beqn
{\rm Re}\hat{\Sigma}_{ij}^{\prime}(M_{i,{\rm tree}}^2)= {\rm Re}\hat{\Sigma}_{ij}^{\prime}(M_{j,{\rm tree}}^2)=0 \quad {\rm for } \quad i \neq j
\eeqn
which then fixes the non-diagonal elements of the wave-function renormalisation matrices such that 
\beqn
\delta Z_{ij}= 2 \frac{{\rm Re}\Sigma_{ij}(M_{j,{\rm tree}}^2) + \delta M_{ij}^2}{M_{j,{\rm tree}}^2-M_{i,{\rm tree}}^2} \quad i \neq j.
\eeqn
An example of the latter is 
\beqn
\delta Z_{h_i^0h_j^0}=2\frac{{\rm Re}\Sigma_{h_i^0h_j^0}(M_{h_j^0}^2)+\delta M_{h_i^0h_j^0}^2}{M_{h_i^0}^2-M_{h_j^0}^2} \quad i \neq j
\eeqn

\section{Renormalisation schemes}
\label{sec:ren_schemes}
The definition of the underlying counterterms involve solving a system of coupled equations which, moreover,  depends crucially  on the choice of the input parameters, for example which set of physical masses or  other observables one chooses as input. In this respect, the wave-function renormalisation constants are somehow  easy to evaluate. They involve a one-to-one relation and are independent from each other. Their expression is independent of the scheme.  

\noi We have also already  specified, see
Section~\ref{subsec:counting_para} how the SM parameters, $g, g^\prime, v
\leftrightarrow M_W, M_Z, e$ that enter also in the NMSSM, are renormalised.

The reconstruction of the counterterms of the 9 underlying parameters of the
Higgs sector in Eq.~\ref{para_count_higgs_2} is more complicated. Indeed, most
of these parameters contribute to more than one Higgs mass (and chargino/neutralino 
mass) or Higgs observable. Apart from the tadpoles, defined from
Eq.~\ref{eq:tadpole_ct}, it is not obvious what the optimal set of the 9 input
parameters should be  in order to reconstruct these underlying parameters. Leaving the
tadpole aside, there remains 6 parameters to determine in the Higgs sector. In
principle, it is possible to use only masses as inputs since the Higgs sector does
furnish 6 different Higgs masses,
$h_{1,2,3}^0, A_{1,2}^0,H^\pm$. Technically, this requires the computation of the
self-energies. Note  that four of these parameters,
$t_\beta, \lambda, \mu, \kappa$,  could also just as well be determined from the
neutralino/chargino sector. However, as we argued in detail in
\cite{Belanger:2016tqb}, one cannot solve for the latter 4 parameters alone
since the chargino/neutralino system involves also the underlying parameters
$M_1,M_2$ (the $U(1)$ and $SU(2)$ gaugino soft masses). The two parameters
$A_\lambda$ and $A_\kappa$ can only be defined in the Higgs sector. The CP-even
and CP-odd masses depend on both
 these parameters, while the charged Higgs mass depends on $A_\lambda$ only.
Because of the intrinsic interdependence of the NMSSM observables related to the
Higgs sector (and the corresponding neutralino/chargino sector), in all
generality we need to consider a system of counterterms for the 8 underlying
parameters which in  a vector notation reads

\begin{equation}
\label{para_count_higgs_neut}
\mathbbm{p}=\left(\underbrace{t_\beta, \lambda,\kappa (m_\k),\mu}_{{\rm in} \; \tilde{\chi} \; {\rm and} \; {\rm Higgs} \; {\rm sectors}}, \underbrace{A_{\lambda}, A_\kappa}_{{\rm Higgs}}, \underbrace{M_1,M_2}_{\tilde{\chi}} \right),
\end{equation}
for which we need to select 8 input parameters with at least two from the chargino/neutralino sector and at least two from the Higgs system. Let us recall the procedure and the most important points that we detailed in \cite{Belanger:2016tqb} for reconstructing all 8 counterterms. Injecting 8 input parameters we have to solve for 
\begin{align}
\begin{pmatrix}
\delta {\tt input}_1 \\
\cdots\\
\cdots \\
\delta {\tt input}_8\\
\end{pmatrix}={\mathcal{P}}_{8, {\rm param.}} \;
\delta\mathbbm{p}+{\mathcal{R}}_{8,\rm residual},
\label{Pnn}
\end{align}
${\mathcal{P}}_{8, {\rm param.}}$ is a $8\times8$ matrix, ${\mathcal{R}}_{n,\rm
residual}$ contains other counterterms, such as gauge
couplings,  that are defined separately. Using the physical mass of one of the
Higgs boson as an input, see Eq.~\ref{eq:input_mhp}, is a possible choice 
in an OS scheme. Not all inputs need to be OS. For an efficient resolution of
Eq.~\ref{Pnn}, {\it i.e.} determining the  $\delta\mathbbm{p}$ vector, one
should break up the connecting matrix ${\mathcal{P}}_{8, {\rm param.}}$ in as
many, possibly smallest rank, matrices as possible
\beqn
\mathcal{P}_{8, {\rm param.}} = \mathcal{P}_{m, {\rm param.}} \oplus \mathcal{P}_{p, {\rm param.}} \oplus \cdots,
\quad m+p+\cdots=8
\eeqn
The choice of the input parameters will determine how one can build the
$\mathcal{P}_{8, {\rm param.}}$ from smaller independent blocks, each choice
will define a renormalisation scheme. It is important to seek a scheme where the
determinant of each sub-matrix $\mathcal{P}_{p, {\rm param.}}$ is not too small ($\delta\mathbbm{p} \propto 1/{\rm Det }\mathcal{P}$) in order not to introduce large coefficient that would lead to large radiative
corrections solely from a bad choice of inputs. We will pursue the comparisons
between different schemes in the applications to  Higgs decays  in Section~\ref{sec:higgs_decays}.

\subsection{Mixed OS-$\overline {\textrm{DR}}$
schemes\label{sec:mixscheme_notation}}
Realising that $t_\beta$ is ubiquitous, it even enters the determination of the
wave-function renormalisation matrices of the Higgses, a practical possibility
is to take a $\overline {\textrm{DR}}$ condition for  $t_\beta$. In this case,
the sectors for the Higgs, the neutralinos and the charginos can be solved
independently. The
counterterms for $\mu$ and the $SU(2)$ gaugino mass term $M_2$ can be extracted
from both chargino masses. Then, the counterterms for the self-interacting singlet coupling $\kappa$ (through  $m_\k$), the $U(1)$
gaugino mass term $M_1$ and the singlet-doublet coupling $\l$ can be obtained
from the masses of the neutralinos that
should be chosen to represent the mainly singlino, bino and higgsino
neutralinos. We have
commented at length about the issue of a {\em knowledge} of the nature (content)
of the neutralinos in \cite{Belanger:2016tqb}. 
As concerns $A_\lambda, A_\kappa$, a possibility is for example to use both
CP-odd Higgs bosons as inputs, $\delta M_{A_i^0}^2={\rm
Re}\Sigma_{A_i^0A_i^0}(M_{A_i^0}^2), i=1,2$. This break-up corresponds to 

\beqn
\mathcal{P}_{8, {\rm param.} }=\underbrace{\mathcal{P}_{1, {\rm
param.}}}_{\overline {\textrm{DR}},t_\beta}\oplus \underbrace{\mathcal{P}_{2,
{\rm param.}}}_{OS, M_{\chi_{1,2}^\pm}}
\oplus \underbrace{\mathcal{P}_{3, {\rm param.}}}_{OS, M_{\chi_{1,2,3}^0}}\oplus
\underbrace{\mathcal{P}_{2, {\rm param.}}}_{OS, M_{A_{1,2}^0}}
\eeqn


\noi The $\overline{\textrm{DR}}$ condition for $t_\beta$ in this scheme  is an extension of the DCPR
scheme\cite{Chankowski:1992er,Dabelstein:1994hb}, used in the context of the
MSSM, to the NMSSM\cite{Ender:2011qh},
\begin{align}
\frac{\delta t_\beta}{t_\beta}=\frac{1}{2}(\delta Z_{H_u}-\delta
Z_{H_d}){|}_\infty,
\end{align}
where $\delta Z_{H_u}$ and $\delta Z_{H_d}$ are the wave function renormalisation constants of the $H_u$ and $H_d$
doublets. The infinity symbol indicates that we only take the divergent part of the expression. $\delta Z_{H_u}$ and
$\delta Z_{H_d}$ are related to the wave function renormalisation constants
$\d Z_{h_i h_i}$ (Eqs.~\ref{eq:Z_S}) through
\beqn
\delta Z_{H_d}&=&\frac{1}{R}\sum_{i,j,k=1}^3\epsilon_{ijk}S_{h,j3}S_{h,k2}\delta Z_{h_ih_i}, \quad
\delta Z_{H_u}=\frac{1}{R}\sum_{i,j,k=1}^3\epsilon_{ijk}S_{h,j1}S_{h,k3}\delta Z_{h_ih_i},\nonumber \\
R&=&-\sum_{i,j,k=1}^3\epsilon_{ijk}S_{h_i1}^2S_{h,j2}^2S_{h,k3}^2,
\eeqn
where $\epsilon_{ijk}$ is the fully antisymmetric rank 3 tensor with $\epsilon_{123}=1$.  

\noi In a $\overline{\textrm{DR}}$ scheme only the divergent part of the countertem is defined \textit{i.e}, any
finite term is set to $0$. Nonetheless, the scheme and the one-loop result is still not fully defined unless one
specifies the renormalisation scale $\bar{\mu}$. The latter is the remnant scale introduced by the regularization
procedure, dimensional reduction. This class of renormalisation schemes will be
denoted as $t_{ijkA_1A_2}$. The letters $i,j,k$ refer to the neutralinos
whose mass has been taken as input. $A_1A_2$  is a reminder that the masses of the two physical pseudoscalars have been used as input also.


This disentagled scheme is rather simple to implement, but can lead to a poor
extraction of $\delta\lambda$, since this parameter is present only in non
diagonal entries of  the neutralino mass matrix. A solution is to take another
Higgs mass as input to get this counterterm, but in this case the neutralino and
the Higgs sectors are no longer disassociated. 

\subsection{Fully on-shell schemes\label{sec:osscheme_notation}}
For these schemes,  the set of eight counterterms, including $t_\beta$, are
obtained from OS conditions based on  inputs taken from physical masses.   As for the
previous scheme, $\delta\mu$ and $\delta M_2$ will be mainly reconstructed from
the two  chargino masses but not fully since there is still some mixing
introduced by $t_\beta$. In addition, it is natural to take the neutralino that
is mainly bino to extract  $\delta M_1$. The parameters $\delta \lambda$ and
$\delta m_\kappa$ can be extracted either from the neutralino or the Higgs
sector. As before, $\delta A_\lambda$ and $\delta A_\kappa$ have to be
extracted from two masses from  the Higgs sector including at least a mostly
singlet Higgs. Finally,  it is much better to obtain  $\delta t_\beta$ from an
additional Higgs input than from the neutralino masses, as was shown in
~\cite{Belanger:2016tqb}. We are  then left with a system of eight equations,
whose inversion will give all these counterterms.

Different classes of such (fully) on-shell renormalisation schemes are possible. For instance, one can take  one example from the general class where the  masses of the two charginos is exploited. This furnishes a system of two equations. On can then have variations on this scheme depending on which source provides the other parameters.

\begin{itemize}
\item One scheme could use the masses of three neutralinos  as well the masses of  both pseudoscalars Higgses and the mass of the charged Higgs. This choice is referred to as  $OS_{ijkA_1A_2H^+}$ where $i,j,k=1...5$ designates the three chosen neutralinos, usually a bino, a singlino and a higgsino. 
\item Another choice could be based on the masses of (only) two neutralinos, both pseudoscalars
Higgses, the charged Higgs and an additional CP-even Higgs different from the
SM-like Higgs. This subclass is
labelled as $OS_{ijh_{\hat{\i}} A_1A_2H^+}$ where $i,j=1...5$ and
$\hat{\i}=1,2,3$. 
\item The third possibility is to use the mass of one neutralino, the bino, and the masses of 5
Higgses, in the obvious notation, $OS_{ih_{\hat{\i}} h_{\hat{\j}} A_1A_2H^+}$
where $i=1...5$ and $\hat{\i},\hat{\j}=1,2,3$.
\end{itemize}

The numerical examples we will consider are based on the $OS_{ijh_{\hat{\i}}
A_1A_2H^+}$ with an optimal choice for the neutralinos. 

\section{Checks on the results and tracking the scheme dependence}
\label{sec:beta_dep}
To check the validity of our numerical results, we perform several tests. The
most powerful test consists in checking the absence of  ultraviolet divergences
on all the  observables that we calculate.  Ultraviolet  divergences appear in
many intermediate steps of the calculations and  get cancelled out by the
counterterms for the underlying parameters and/or among many diagrams.  The
ultraviolet divergences are encoded in the parameter $C_{UV}$ defined in
dimensional reduction as $C_{UV}=2/\epsilon -\gamma_E+\ln(4\pi)$  where
$\epsilon=4-d$ ($d$ being the number of dimensions) and $\gamma_E$ is the Euler
constant. We systematically check that  the numerical results, for one-loop
corrections to masses or to decay processes, are independent of $C_{UV}$ by
varying this parameter from 0 to $10^7$. 
 We require that the numerical results agree for at least seven digits
(\texttt{SloopS} uses double precision).  Thus we ensure that physical processes
are finite for all renormalisation schemes.

Finally, in schemes where at least one parameter is taken to be $\drbar$, a
dependence on the renormalisation scale $\bar\mu$ remains, this can be used to
quantify the scale dependence.  

\subsection{The $\beta$ functions and the scale dependence
\label{subsec:beta_dep}}
In order to gain a qualitative understanding of the differences in the results
for the one-loop corrections to the decay processes we have studied in
different schemes, it is interesting to recall some simple arguments related to
the counterterms. The infinite ($C_{UV}$) part of any counterterm is obviously
the same regardless of the renormalisation scheme, this is one of  the reasons
our checks show finiteness for the calculated observable for all schemes. The
finite part of the counterterm is however scheme dependent. The difference in
this finite part for a particular parameter can be large between two schemes. This difference may get
amplified in the computation of an observable that depends strongly on this particular parameter. This parametric dependence that can be derived from the study of the observable at tree-level is therefore an important ingredient also. 
\noi Take a parameter $\mathbbm{p}_i$, its counterterm, within some scheme
$Q_{\mathbbm{p}_i}$ reads 
 \beqn
 \label{delta_os}
 \delta \ppi/\ppi= \beta_{\ppi} (C_{UV}+\ln (\bar{\mu}^2/Q_{\ppi}^2))
 \eeqn
 $\bar{\mu}$ is the scale introduced by dimensional reduction.
$\beta_{\ppi}$ as defined here is the one-loop $\beta$ constant for the parameter $\ppi$. It  is scheme
independent.  $Q_{\ppi}^2$ encodes the scheme dependence. In this notation, different schemes
correspond to different values of $Q_{\ppi}^2$. $Q_{\ppi}^2$ is  the square of some mass scale
which represents both external momenta (corresponding to the choice of the
subtraction points) and internal masses typical of two-point functions. All our
renormalisation schemes are based on two-point functions. If $Q_{\ppi}^2$ is
dominated by a mass  $m_P$
much larger than all other scales in the problem then $Q_{\ppi} \sim m_P$. Note however that these two-point functions can also
involve a non-log constant terms, in our definition these non-log terms are lumped into
$Q_{\ppi}$. Within the same scheme, 
$Q_{\ppi}^2$, the difference in the value of the counterterm due to a change in the regularisation scale $\bar\mu$ is a measure of $\beta_{\ppi}$
\beqn
\Delta_{\mu_2-\mu_1} \delta \ppi/\ppi= \delta \ppi(\bar{\mu}_2)/\ppi-\delta
\ppi(\bar{\mu}_1)/\ppi=\beta_{\ppi} \ln (\bar{\mu}_2^2/\bar{\mu}_1^2)
\eeqn
In our code this is how we determine $\beta_{\ppi}$ numerically.  In our
numerical analysis in Sec.~\ref{sec:higgs_decays} these $\beta$ constants have been checked against
the $\beta$ functions given in \cite{Ellwanger:2009dp}. Perfect agreement has
been found when specialising to the one-loop result. This is a non-trivial check on our renormalisation procedure. For the so-called
$\drbar$ scheme, we set the corresponding counterterm to
\beqn
 \delta^{\drbar} \ppi/\ppi= \beta_{\ppi} C_{UV}, 
 \eeqn
which in effect corresponds to choosing a scale $Q_{\ppi} =\bar\mu$. \\
\noi The one-loop correction to an observable involves calculating all the
virtual two-point, three-point, ...n-point functions which are specific for a
given
amplitude (regardless of the scheme) and then including the counterterms for all
parameters on which the amplitude ${\mathcal O}$ depends on, to obtain a finite
result. The dependence of the amplitude on a specific parameter
$\ppi$, the parametric dependence alluded to earlier, is obviously also very
important. If at tree-level we slightly change the
value of the parameter $\ppi$ by an amount $\delta \ppi$, we define the percentage change on the observable as \beqn
\label{def_kapi}
\frac{\delta {\mathcal O}}{{\mathcal O}}= \sum_i \kappa_{\ppi} \frac{\delta\ppi}{\ppi} 
\eeqn
The sum is over all the {\it independent} parameters of the model. If an observable is
independent of a particular parameter $\ppi$ then the corresponding $\kappa_{\ppi}$ is $\kappa_{\ppi} =0$\footnote{Strictly speaking the sum applies to each Lorentz structure
and/or helicity amplitude.}. In this definition of the parametric dependence we are assuming small, infinitesimal, $\delta \ppi$ as is generally the case when $\delta \ppi$ stand for one-loop counterterms or else that the dependence in the parameter $\ppi$ is linear. With this proviso and to make the discussion simple, 
if all counterterms are defined on-shell, or at some subtraction point according
to Eq.~\ref{delta_os}, then the virtual corrections  for the amplitude can be
written in a very compact way as 
\beqn
\label{change-os}
\delta {\mathcal O}^{{\rm OS}}/{\mathcal O} = \sum_i \beta _{\ppi} \kappa_{\ppi}
\ln (Q_\Delta^2/Q_{\ppi}^2) 
\eeqn
The correction can be large  if some $\beta$ constants  are large. The corrections could be  
also amplified if the parametric dependence on the parameter $\kappa_{\ppi}$ is
large and/or  if there is a large difference between some subtraction scale in
defining a particular parameter namely $Q_{\ppi}^2$ and the scale that defines the observable
$Q_\Delta$. $Q_\Delta$ can have contributions not only from two-point functions
but also from n-point functions  which we have lumped in its definition. In
particular our parameterisation of   $Q_\Delta$ can take into account  non
single logarithms (dilogarithms, $\cdots$). In particular some one loop dynamics may entail large genuine corrections which are then translated here as large values of $Q_\Delta$. \\

\noi In a scheme where \underline{all} counterterms are defined \`a la $\drbar$,
 \beqn
 \label{change-drbar}
\delta {\mathcal O}^{\drbar}/{\mathcal O} = \ln (Q_\Delta^2/\bar{\mu}^2) \sum_i
\beta _{\ppi} \kappa_{\ppi}. \eeqn
There is now a scale dependence which quantifies the uncertainty in the one-loop calculation. The correction is minimised for $\bar \mu \sim Q_\Delta$. Again if the virtual corrections are dominated by single logs with an argument corresponding to the largest scale/mass of the process, the corrections are minimised for $\bar \mu$ corresponding to this highest scale. \\

\noi In a mixed-scheme like the one we have taken with $\mathbbm{p}_0=t_\beta$
and with $Q_ {\mathbbm{p}_0}$ the effective scale that defines the OS definition
of $t_\beta$, the result for the correction to the same observable can be
written as\footnote{ In a scheme where all parameters are defined on-shell
according to Eq.~\ref{delta_os}, the numerical extraction of the $\beta$
constants through the $\mu$ variation is quite simple, it relies on the
combination $(C_{UV}+\ln \bar\mu^2)$. In a mixed scheme like the one  where $t_\beta$ is $\drbar$ and the other underlying parameters are
reconstructed from solving a coupled system based on on-shell quantities through
masses as input, there may be a mismatch between the coefficient
multiplying $C_{UV}$ and $\ln\bar\mu^2$.}
\beqn
\delta {\mathcal O}^{\rm{mixed}}/{\mathcal O} =\delta {\mathcal O}^{{\rm
OS}}/{\mathcal O}  + \beta_{\mathbbm{p}_0}\kappa_{\mathbbm{p}_0} \ln
(Q_{\mathbbm{p}_0}^2/\bar{\mu}^2)  
\eeqn

The $\beta$ functions can also be derived from an analysis of the
renormalisation group. The system of coupled equations at 2-loop  is given in
\cite{Ellwanger:2009dp}. Specialising to one-loop and keeping only the
dominant Yukawa coupling contributions, the system is rather simple and can help
understand some features of the full one-loop calculation. With $h_t$ the
top-Yukawa coupling, the dominant contributions to the running of the underlying
couplings of the NMSSM can be cast as  
\beqn
\label{approx-beta}
 \frac{1}{h_t^2} \frac{{\rm d} h_t^2}{ {\rm d}\tau}=\frac{1}{A_t} \frac{{\rm d} A_t}{ {\rm d}\tau}=\frac{2}{\lambda^2} \frac{{\rm d} \lambda^2}{ {\rm d}\tau}= \frac{2}{\mu^2} \frac{{\rm d} \mu^2}{ {\rm d}\tau}= 6 h_t^2 \quad {\rm} {\rm and} \quad
\frac{1}{A_\lambda} \frac{{\rm d} A_\lambda}{ {\rm d}\tau}=3 h_t^2 \frac{A_t}{A_\lambda},
\eeqn
with $\tau=\ln \bar\mu^2/16\pi^2$. What this shows is that if $A_t \gg A_\l$ then 
$\beta_{A_\l}$ can be quite large. Remember that a large $A_t$ is needed for
inducing a large one-loop correction to the MSSM-like CP-even Higgs (in the MSSM
limit). Note that $A_{\kappa}$ and $s=\mu/\l$ do not have top-Yukawa enhanced running.

\subsection{Infrared divergences}
Many of the Higgs processes we will study at the one-loop level, will give
rise to infrared divergences when electrically charged particles are involved in
the external legs. The treatment of these divergences requires the computation
of real photon  emission as described in ~\cite{Belanger:2016tqb}. In a
nutshell, for these $1 \to 2$ decays it is sufficient to take an infinitesimally
small photon mass as a regulator. For more details see ~\cite{Belanger:2016tqb}.

\section{The benchmark points}
\label{sec:benchmark}
\begin{table}[h]
\begin{center}
\noindent\makebox[\linewidth]{\rule{0.8\paperwidth}{0.4pt}}
\caption{\label{tab:bench2} Parameters for the  benchmark point A (in GeV
for all dimensionful parameters).  $Q_{\rm susy}$ is calculated as $
Q_{\rm susy}=\sqrt{M_{\tilde{t}_1} M_{\tilde{t}_2}}=1117.25$ GeV. The derived
values for the tree-level masses of all Higgses, charginos and neutralinos are
also given. For this benchmark, the top mass, crucial for the computation of the Higgs mass in
particular, is taken as $M_t=175$ GeV.}
\begin{tabular}{|cc|cc|cc|cc|cc|}
\hline
$M_1$&700&$\lambda$&0.1&$A_\kappa$&0&$m_{\tilde{Q}_3}$&1740&$m_{\tilde{D},\tilde{U}_{1,2}}$&1000\\
$M_2$&1000&$m_\kappa${\tiny ($\kappa$)}&120{\tiny ($0.1$)}&$A_t$&4000&$m_{\tilde{U}_3}$&800&$m_{\tilde{L}_3}$&1000\\
$M_3$&1000&$\mu$&120&$A_b$&1000&$m_{\tilde{D}_3}$&1000&$m_{\tilde{l}_{3}}
$&1000\\
$t_\beta$&10&$A_\lambda$&150&$A_l$&1000&$m_{\tilde{Q}_{1,2}}$&1000&$m_{\tilde{L},\tilde{l}_{1,2}}$&1000\\
 \hline
\end{tabular}
\end{center}
{\small
$(M_{\chi^+_1},M_{\chi^+_2};M_{\neuto},M_{\chi^0_2},M_{\neuth},M_{\neutf}
,M_{\chi^0_5})=(117.95,1006.61;112.77,123.80,241.57,702.82,1006.64) $}\\
{\small $(M_{H^\pm};M_{A_1^0}, M_{A_2^0};M_{h_1^0},M_{h_2^0},M_{h_3^0})=(577.33;
12.64,572.06; 88.47,240.07,572.48)$}\\
The one-loop corrected SM-like Higgs mass is calculated to be 125.45 GeV in the
${\rm OS}_{34h_2A_1A_2H^+}$ and 126.47 GeV in the $\drbar$ scheme  with a scale
at $Q_{\rm susy}$.\\
\noindent\makebox[\linewidth]{\rule{0.8\paperwidth}{0.4pt}}
\end{table}

We will concentrate on two quite distinct scenarios of the NMSSM. 
The first scenario, Point A, is chosen with a very small value of the mixing parameter $\lambda$ in order to study the  MSSM limit. Point B has a much larger value of $\l$ exhibiting large mixing between the singlet and doublet components. \\

\noi Point A  is defined through practically the same parameters that we chose to set the benchmark Point 3 in  our previous work~\cite{Belanger:2016tqb} on the renormalisation of the chargino/neutralino sector. The defining parameters of Point A are listed in Table~\ref{tab:bench2}.
An alert reader would have noticed that the difference between Point 3 in
\cite{Belanger:2016tqb} and Point A  is that the values of the stop masses were
modified to ensure that the one-loop corrected mass for the SM-like Higgs be compatible with the
value observed at the LHC. Because of the small value of $\lambda$, we are in the  MSSM limit which requires 
to take a fairly large value of the trilinear  parameter, $A_t$,  in the stop sector. Observe
that the value of $A_t$ is very large compared to $A_\l$ with $A_t/A_\l\sim 27$.
This will have important side effects apart from giving large corrections to
the lightest CP-even neutral Higgs. We have also listed the value of the
one-loop corrected mass for the SM-like Higgs. We see that it is compatible with
the mass of the Higgs discovered at the LHC. Table~\ref{tab:bench2} gives the value of this mass in
two schemes. The scheme difference is within 1 GeV. For more details on the
correction to the Higgs masses and tuned comparisons with other calculations we
refer to \cite{bizouard:tel-01447488}. \\

\noi For the second benchmark, we borrowed parameters very similar to Point TP4
in \cite{bizouard:tel-01447488}. Benchmark B is defined in 
Table~\ref{tab:benchTP4}.
\begin{table}[h]
\begin{center}
\noindent\makebox[\linewidth]{\rule{0.8\paperwidth}{0.4pt}}
\caption{\label{tab:benchTP4} Parameters for the  benchmark  Point B (in GeV
for all dimensionful parameters).   $Q_{\rm susy}$ is calculated as $
Q_{\rm susy}=\sqrt{M_{\tilde{t}_1} M_{\tilde{t}_2}}=753.55$ GeV. The derived
values for the tree-level masses of all Higgses, charginos and neutralinos are
also given. }
\begin{tabular}{|cc|cc|cc|cc|cc|}
\hline
$M_1$&120&$\lambda$&0.67&$A_\kappa$&0&$m_{\tilde{Q}_3}$&750&$m_{\tilde{D},
\tilde {U}_{1,2}}$&1500\\
$M_2$&300&$m_\kappa${\tiny ($\kappa$)}&59.7{\tiny ($0.2$)}&$A_t$&1000&$m_{\tilde{U}_3}$&750&$m_{\tilde{L}_3}
$&1500\\
$M_3$&1500&$\mu$&200&$A_b$&1000&$m_{\tilde{D}_3}$&1500&$m_{\tilde{l}_{3}}
$&1500\\
$t_\beta$&1.92&$A_\lambda$&405&$A_l$&1000&$m_{\tilde{Q}_{1,2}}$&1500&$m_{\tilde{
L } ,\tilde{l}_{1,2}}$&1500\\
 \hline
\end{tabular}
\end{center}
$(M_{\chi^+_1},M_{\chi^+_2};M_{\neuto},M_{\chi^0_2},M_{\neuth},M_{\neutf},
M_{\chi^0_5})=(159.63,342.70;89.99,143.90,196.18,235.64,344.98
)
$\\
$(M_{H^\pm};M_{A_1^0}, M_{A_2^0};M_{h_1^0},M_{h_2^0},M_{h_3^0})=(469.09;
111.59,481.56; 102.92,142.84,479.00)$\\
The one-loop corrected SM-like Higgs mass is calculated to be 124.44 GeV in the ${\rm OS}_{12h_2A_1A_2H^+}$ and 121.62 GeV in the $\drbar$ scheme with a scale at $Q_{\rm susy}$. To calculate the Higgs mass we have taken a running top mass at the scale $Q_{\rm susy} =
\sqrt{M_{\tilde{t}_1}M_{\tilde{t}_2}}$ with $M_t=146.94$ GeV. 
\\
\noindent\makebox[\linewidth]{\rule{0.8\paperwidth}{0.4pt}}
\end{table}

\noi One notable difference between the two benchmark points is the value of
$\lambda$ (6 times larger for Point B), such that for Point B, $\Lambda_v >
M_Z$. As a consequence, for point B the tree-level value for the mass of the
SM-like Higgs is larger than $M_Z$ which is not the case for Point A. This is
the reason why for Point B the value of $\tb$ is $\sim 2$ and more importantly
$A_t/A_\lambda \sim 2.5$ only. Still, one needs radiative corrections to lift
the mass of the SM-like Higgs  from $103$ GeV at tree-level to about $125$ GeV.  \\

\noi Because we will study Higgs decays either to other Higgses or to
neutralinos and charginos, the field content (in terms of the current, unmixed,
fields) is very important. The field content or the purity of the physical
fields, at tree-level, is given in Table~\ref{tab:compo}. If we arrange
the physical fields in terms of their dominant component, then for 
\beqn
\label{dominanceA}
& & {\rm Point \; A} \nonumber  \\
&
&(\neuto,\neutt,\neuth,\neutf,\neutfi;\charg,\chargt
) \sim 
(\tilde{H}^0,\tilde{H}^0,\tilde{S}^0,\tilde{B}^0,\tilde{W}^0_3;\tilde{H}^+,\tilde{W}^+) \nonumber \\ & & (h_1^0,h_2^0,h^0_3;A_1^0, A_2^0 )\sim (h_u^0,h_s^0,h_d^0;a_s^0,a_d^0).
\eeqn
\noi For Point A  the states have a very high degree of purity. Given the fact
that $\neuto$ is mostly higgsino, $\neuth$ is mostly singlino and $\neutf$
mostly bino, this justifies to use the $t_{134A_1A_2}$ mixed OS-$\drbar$
renormalisation scheme, following the notation of
Sec.~\ref{sec:mixscheme_notation} and the $OS_{34 h_2A_1A_2 H+}$ renormalisation
scheme, following the notation of Sec.~\ref{sec:osscheme_notation}, to compute
the one-loop corrections. Indeed, as discussed in~\cite{Belanger:2016tqb}, in
choosing the input masses, on should preferably include the bino and singlino
from the neutralino sector, and the higgsino when a third neutralino is to be
used.\\

\noi For point B, there is strong mixing and only the lightest physical pseudoscalar field  can be described as pure, nonetheless we can write the dominant components 
\beqn
\label{dominanceB}
& &{\rm Point \; B} \nonumber \\
& &(\neuto,\neutt,\neuth,\neutf,\neutfi;\charg,\chargt)  \sim 
(\tilde{B}^0,\tilde{S}^0,\tilde{H}^0,\tilde{H}^0,\tilde{W}^0_3;\tilde{H}^+,\tilde{W}^+) \nonumber \\ 
& & (h_1^0,h_2^0,h^0_3;A_1^0, A_2^0) \sim (h_u^0,h_s^0,h_d^0;a_s^0,a_d^0).
\eeqn
We will therefore use the $t_{123A_1A_2}$ and $OS_{12 h_2A_1A_2H+}$ renormalisation
schemes to compute the radiative corrections.
\begin{table}[htb]
\noindent\makebox[\linewidth]{\rule{0.8\paperwidth}{0.4pt}}
\begin{center}\caption{Components the  mass eigenstates for  benchmark
points A and B. The dominant component
is highlighted. \label{tab:compo}}
\end{center}
\begin{minipage}{0.495\textwidth}
\begin{center}
\begin{tabular}{cccccc}
\hline
&&Point A &Point B \\
\hline $h_1^0$&$h_d^0$& 1.1\% & 22.5\% \\
                       &$h_u^0$& \textbf{98.6\%} & \textbf{67.4\%} \\
                       &$h_s^0$&  0.3\%& 10.1\%\\    
                       &&& \\
         $h_2^0$&$h_d^0$& 0.1\% & 0.\% \\
                       &$h_u^0$&  0.3\%& 12.5\% \\
                       &$h_s^0$&  \textbf{99.6\%}& \textbf{87.5\%} \\   
                       &&& \\   
         $h_3^0$&$h_d^0$& \textbf{98.8\%}  &\textbf{77.5\%}  \\
                       &$h_u^0$&  1.1\%&  19.7\%\\
                       &$h_s^0$&  0.1\% & 2.8\% \\    
                       &&& \\
         $A_1^0$&$a_d^0$& 0\%& 1.8\%  \\
		       &$a_u^0$& 0\% & 0.5\% \\
		       &$a_s^0$&\textbf{ 100\%} & \textbf{97.7\%} \\
                       &&& \\         
          $A_2^0$&$a_d^0$& \textbf{99.0\%} & \textbf{76.9\%} 
  \\
		       &$a_u^0$& 1.0\% & 20.8\% \\
		       &$a_s^0$&  0.0\%& 2.3\% \\                 
\hline
\end{tabular}
\end{center}
\end{minipage}
\begin{minipage}{0.495\textwidth}
\begin{center}
\begin{tabular}{cccc}
\hline
&&Point A &Point B \\
\hline
$\tilde{\chi}_1^0$&$\tilde{B}^0$&- &{\bf 56.6\%} \\
                           &$\tilde{W}^0$&- &32.3\% \\
			 &$\tilde{h}^0$&{\bf 98.4\%} &10.3\% \\
			 &$\tilde{S}^0$& 0.77\%&0.8\% \\\hline
$\tilde{\chi}_2^0$&$\tilde{B}^0$& -&4.0\%\\
			 &$\tilde{W}^0$& -&2.6\% \\
			 &$\tilde{h}^0$& {\bf 99.5\%}&19.3\% \\
			 &$\tilde{S}^0$&  -&{\bf 74.0\%} \\\hline
$\tilde{\chi}_3^0$&$\tilde{B}^0$& -&10.1\%\\
			 &$\tilde{W}^0$&- & - \\
			 &$\tilde{h}^0$& 0.9\%&{\bf 78.9\%} \\
			 &$\tilde{S}^0$& {\bf 99.1\%} &11.0\% \\\hline
$\tilde{\chi}_4^0$&$\tilde{B}^0$& {\bf 99.6\%}&18.1\% \\
			 &$\tilde{W}^0$& -&12.3\% \\
			 &$\tilde{h}^0$&- &{\bf 55.8\%}\\
			 &$\tilde{S}^0$&- &13.7\%\\\hline
$\tilde{\chi}_5^0$&$\tilde{B}^0$&- & 11.2\% \\
			 &$\tilde{W}^0$& {\bf 99.3\%}& {\bf 52.8\% }\\
			 &$\tilde{h}^0$& 0.69\%& 35.7\% \\
			 &$\tilde{S}^0$&- & 0.4\%\\
\hline
\end{tabular}
\end{center}
\end{minipage}
\noindent\makebox[\linewidth]{\rule{0.8\paperwidth}{0.4pt}}
\end{table}

\section{Higgs decays
\label{sec:higgs_decays}}
As an application to our set-up for the renormalisation of the Higgs sector and its implementation in {\tt SloopS} we consider Higgs decays. This covers  decays of Higgses into neutralinos and charginos, final states with a single gauge boson as well as decays into lighter Higgses. These channels also serve to test the most critical aspects of the renormalisation of the Higgs sector in the NMSSM. We have not computed decays involving sfermions
since the sfermion sector does not introduce much novelty compared to the MSSM, nor did we consider here decays into SM fermions and pairs of gauge bosons. Note that we have computed decays of the neutral Higgs scalars to $\gamma 
\gamma$ and $Z\gamma$ in an earlier publication\cite{Belanger:2014roa}, however these loop induced decays do not require renormalisation. \\

\noi The importance of the radiative corrections and the choice of the
renormalisation scheme underline the importance of studying the parametric
dependence. To gain an understanding, at least qualitatively, of the results of
some of the radiative corrections for the most prominent decays of the Higgses,
we will first show, for both Points A and B,  how the value of the corresponding
{\em tree-level}  partial width changes when one of the underlying
parameters is modified  around each one of  the reference points that define the model. As discussed in section~\ref{subsec:beta_dep} this will 
give us an insight on the parametric dependence and an approximate extraction of
the coefficients $\kappa_{\ppi}$ (see Eq.~\ref{def_kapi}) when specialising to
small variations. 
We will in fact only show the variations of the square of the coupling involved
in the decay. This quantity  represents the square of the amplitude for the 
partial width, leaving  the phase space factor out. The rationale for doing this
is that a variation of the underlying parameters changes also the values of the
masses which in turn change the phase space and hence introduce another source
of change in the partial width. In the renormalisation process some of the
masses of the particles taking part in the process are taken as input parameters
with a  value fixed at all orders. \\

\noi Before giving the results for the full one-loop corrections to the decays, we will first extract the universal $\beta_{\ppi}$ for each parameter $\ppi$ as explained in section~\ref{subsec:beta_dep}. For each scheme we will also give the value of the finite term (see Eq.~\ref{delta_os}) of the counterterm to the parameter $\ppi$
\beqn
\label{eq:finite_ct}
\frac{\delta \ppi}{\ppi}\bigg\rvert_{{{\rm finite}}}= 
\beta_{\ppi} \ln (\bar{\mu}^2/Q_{\ppi}^2)
\eeqn
evaluated at a value of  $\bar \mu$ which we will specify. For later reference, observe that a large value of $\beta_{\ppi}$ will most certainly entail a large value for the finite part of the corresponding counterterm, unless $\bar{\mu}^2 \sim Q_{\ppi}^2$.  
Remembering our  discussion in section~\ref{subsec:beta_dep} (see also
Eqs.~\ref{change-os} and \ref{change-drbar}), the $\beta_{\ppi}$ and the finite
part of the counterterms in a given scheme, together with what we will have
learnt about the parametric dependence, will help gain some understanding of
the results of the full one-loop corrections, the scheme dependence. One could
also learn   whether there may be large genuine corrections that stem from the 
two and three-point functions or even from the real corrections
(bremmstrahlung). 
 
\subsection{Point A}
For this point we will compute the full electroweak
corrections to the partial decay
widths of CP-even, CP-odd and charged Higgs into other Higgses and
supersymmetric particles. For the latter, only
neutralinos and charginos are kinematically accessible. We will only include  the
channels for which the branching ratio is above 1\% as they are the only
potentially relevant ones. Note that 
the lightest CP-even and CP-odd Higgs decay only into SM particles and that  the
components of the heavy doublet Higgs ($h_3^0,A_2^0$)  also decay mainly
into SM particles, in particular $b\bar{b}$.
Only the  singlet $h_2^0$ decays dominantly in the pair of singlets $A_1^0A_1^0$. The partial widths
of all the channels considered are of the same order at tree-level, about $10^{-2}\;
{\rm GeV}$.

\subsubsection{Tree-level. Parameter dependence on some couplings in Higgs decays}
\begin{figure}[hbt!]
\begin{center}
\caption{Parameter dependence of (the square) of the couplings that enter
some important decays which we will study at one-loop. We look at the variation
in the parameters $\tb, \l, m_\kappa, A_\kappa, A_\lambda$ and $\mu$. Plotted is
the percentage variation measured from the reference point, defined in
Table~\ref{tab:bench2}. We allow variations of $\pm 20\%$ for these parameters
apart from $A_\kappa$ whose reference value, $A_\kappa=0$, is varied smoothly up
to $40$ GeV. The solid (blue) lines represents the $h_3^0 h_2^0 h_1^0$
coupling, the dash-dotted-dotted-dotted (purple) the $A_2^0 Z^0 h_2^0$ 
coupling, the dotted (green) lines  the $h_3^0 \neuto \neuth$ coupling, the
dashed (red) lines the $h_2^0 A_1^0 A_1^0$ coupling, the dash-dotted
(turquoise) lines the $A_2^0 \to \neuto \neuto$ and the long-dashed (gray) 
lines the $A_2^0 \charg \chargm$ coupling.}
\label{fig:pointA-paradep}
\includegraphics[scale=0.8,trim=0 0 0 0,clip=true]{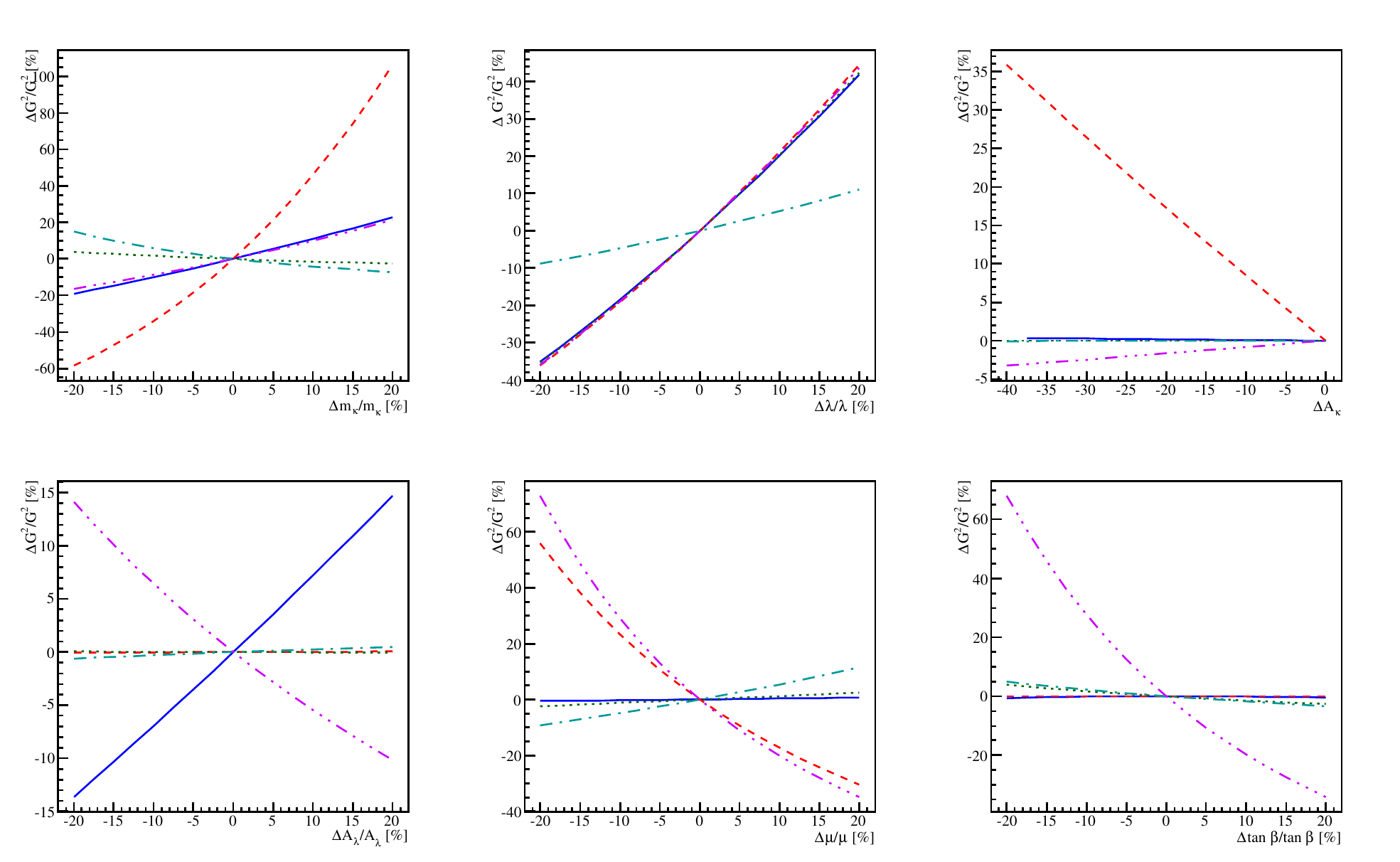}
\end{center}
\end{figure}

As promised, and before going to the loop results, we first look at the parametric dependence of some of the decays we will study. The analysis of the parametric dependence relies on the tree-level behaviour of the observables as the underlying parameters are varied. Results of these variations are shown in Fig.~\ref{fig:pointA-paradep}. The first general observation is the  smooth and  almost linear dependence on all the (independent) parameters, for all the coupling across the whole range of the variations, $\pm 20\%$. This can be easily understood if we recall that this point is characterised by very small mixing $\lambda$  where the physical states have a  high degree of purity, whereby the Higgs states $h_2^0, A_1^0$ and the neutralino $\chi_3^0$ are essentially singlet states. With this small $\l$ scenario, it is instructive to subdivide these decays into three  classes of decays
\begin{itemize}
\item[{\it i)}] All particles involved in the decay are predominantly singlets, $h_2^0\rightarrow A_1^0A_1^0$.
\item[{\it ii)}] None of the particle taking part in the decay is singlet-like with characteristics close to the MSSM and with very little dependence on $\l$. Two  examples are shown in Fig.~\ref{fig:pointA-paradep},  $A_2^0\rightarrow \tilde{\chi}_1^+\tilde{\chi}_1^-$ and $A_2^0 \to \neuto \neuto$. In the one-loop calculation we will also consider $h_3^0\rightarrow \tilde{\chi}_1^+\tilde{\chi}_1^-$.
\item[{\it iii)}] Decays involving both singlets and MSSM like particles.
$h_3^0\rightarrow  h_2^0 h_1^0, A_2^0\rightarrow  Z^0 h_2^0$ (and its $SU(2)$
equivalent  $H^+ \to W^+ h_2^0$) and $h_3^0 \to \neuto \neuth$. These decays
are therefore sensitive to the mixing parameters due to the addition of the
singlet in the NMSSM. For most decays this mixing parameter is
essentially $\l$ while for $h_3^0 h_2^0 h_1^0$ and $A_2^0\rightarrow  Z^0 h_2^0$ $A_\l$ is also crucial.
\end{itemize}
We first look at the decays which involve only singlets and then none of them.

\begin{itemize}
\item $h_2^0\rightarrow A_1^0A_1^0$\\
Similar to the self-coupling of the three CP-even neutral Higgs singlets in Eq.~\ref{hhhlamb}, with $A_\kappa=0$ (at tree-level), the  $h_2^0 A_1^0A_1^0$ interaction stems 
from the term $\kappa^2 S^4$  of the Higgs potential, Eq.~\ref{HiggsPot}. This trilinear coupling is controlled by $\kappa^2 s \propto m_\k^2/s \propto \lambda/\mu \; m_\k^2$. The relative variation of the square of the coupling
\beqn
\label{varia_h2a1a1_A}
\frac{\Delta G^2_{h_2^0 A_1^0A_1^0}}{G^2_{h_2^0 A_1^0A_1^0}} \sim 2
\bigg(\frac{\Delta \lambda}{\lambda} +2 \frac{\Delta m_\k}{m_\k}-\frac{\Delta
\mu}{\mu}\bigg)
\eeqn 
 is well rendered by this simple observation and corresponds very well to the variations shown in Fig.~\ref{fig:pointA-paradep}. The small $\Delta A_\k$ variation in Fig.~\ref{fig:pointA-paradep} can be explained similarly from the Higgs potential. Considering that this coupling is solely within the singlet sector, it is important to stress that the $\l$ dependence here is due to our choice of $m_\k$ (rather than $\kappa$),$\l,\mu$ as independent parameters.

\item $A_2^0\rightarrow \tilde{\chi}_1^0\tilde{\chi}_1^0$ and $A_2^0\rightarrow \tilde{\chi}_1^+\tilde{\chi}_1^-$ \\ 
$A_2^0\rightarrow \tilde{\chi}_1^+\tilde{\chi}_1^-$ is  a MSSM-like decay with a
very small dependence on the mixing  $\l$ scenario and practically independent
of all other parameters. $h_3^0\rightarrow \tilde{\chi}_1^+\tilde{\chi}_1^-$
shares these same features. $A_2^0\rightarrow \tilde{\chi}_1^0\tilde{\chi}_1^0$
is quite similar with however some small dependence that creeps in from the
mixing between the neutralinos. Still, as 
seen from Fig.~\ref{fig:pointA-paradep} the $\l$ dependence is four times
smaller, $\Delta \l/2\l,$ as compared to the situation when a singlet state is
involved in the decay.  One can also note a  small dependence on $\mu$ (recall
that the lightest neutralino is higgsino-like) as well as some $\tb$
dependence. Overall the parametric dependence for the decay $A_2^0\rightarrow
\tilde{\chi}_1^0\tilde{\chi}_1^0$ can be approximated as 
\beqn
\label{a2n1n1}
\frac{\Delta G^2_{A_2^0
\tilde{\chi}_1^0\tilde{\chi}_1^0}}{G^2_{A_2^0\tilde{\chi}_1^0\tilde{\chi}_1^0}}
\sim  \frac{1}{2} \bigg( \frac{\Delta \l}{\l}  + \frac{\Delta \mu}{\mu}
-\frac{\Delta m_\k}{m_\k} -\frac{1}{2}  \frac{\Delta \tb}{\tb} \bigg)
\eeqn
\end{itemize}

We now turn to the decays which involve a mixture of singlets/doublets states. 

\begin{itemize}
\item $h_3^0 \to h_1^0 h_2^0$. \\
This decay is triggered from the coupling  $h_u^0 h_d^0 h_s^0$ whose strength is controlled by $\l (A_\l + 2 m_\kappa)$, see the term $(3,1,2)$ in 
Eq.~\ref{hhhlamb}.  With the values of $m_\k$ and $A_\l$, this dependence gives a relative variation 
\beqn
\label{varia_h1h2h3_A}
\frac{\Delta G^2_{h_3^0 h_1^0 h_2^0}}{G^2_{h_3^0 h_1^0 h_2^0}}&\sim& 2
\bigg(\frac{\Delta \l}{\l}+\frac{A_\l}{A_\l+2 m_\kappa} \frac{\Delta
A_\l}{A_\l}+ \frac{2m_k}{A_\l+2 m_\kappa} \frac{\Delta
m_\k}{m_\k}\bigg)\nonumber \\
&\sim& 2 \bigg(\frac{\Delta \l}{\l}+0.38
\frac{\Delta A_\l}{A_\l}+0.6 \frac{\Delta m_\k}{m_\k}\bigg)
\eeqn
which is extremely well exhibited in Fig.~\ref{fig:pointA-paradep}

\item $h_3^0 \to \neuto \neuth$.\\
This coupling is practically independent of the dimensionful parameters, it is
not generated in the Higgs potential. The coupling is essentially dependent only
on  the mixing $\l$, with a very small $\tb$ dependence,  as confirmed by 
Fig.~\ref{fig:pointA-paradep}. The variation, for the square of the coupling, 
can be parameterised as  
\beqn
\label{varia_h1n1n3_A}
\frac{\Delta G^2_{h_3^0 \neuto \neuth}}{G^2_{h_3^0 \neuto \neuth}}\sim 2
\frac{\Delta \l}{\l}
\eeqn

\item $A_2^0 \to h_2^0 Z^0$.  \\
The coupling responsible for this decay  derives in part from $h_s^0 a_d^0 Z^0$,
therefore the singlet-doublet mixing is an essential ingredient. It is not a
surprise that the $\l$ dependence is the same as with all other decays in this
class. Like with the Higgs self-couplings the $A_\l$ mixing is not negligible as
well as the $\tb$ and $\mu$ dependence which enter through the mixing in the
CP-odd and CP-even Higgs sectors. Apart from the $\l$ dependence, deriving an
analytical formula for the relative variation is difficult. Approximately we
get,  
\beqn
\label{varia_a2h2Z_A}
\frac{\Delta G^2_{Z A_2^0  h_2^0}}{G^2_{Z A_2^0  h_2^0}} \sim 
2 \bigg(
\frac{\Delta \l}{\l} -0.3 \frac{\Delta A_\l}{A_\l}-\frac{\Delta \tb}{\tb}+0.4 \frac{\Delta m_\k}{m_\k} -1.2 \frac{\Delta\mu}{\mu}\bigg).
\eeqn
This is in good agreement with the behaviour seen
in Fig.~\ref{fig:pointA-paradep}. For later reference, observe also that it is
this coupling and the $h_1^0 h_2^0 h_3^0$ that are most sensitive to a variation
in $A_\l$, albeit with opposite trends.

\end{itemize}

\subsubsection{Point A: Finite part of the counterterms and their $\beta$ constant}
The  $\beta_{\ppi}$ constants that we extract numerically (see section~\ref{subsec:beta_dep}) are given  in units of $10^{-3}$,
\begin{align}
 \beta_{\mu} &= 5.70\\
 \beta_{t_\beta} &= 8.44 \\
 \beta_\lambda &= 5.83\\
  \beta_{m_\k}&= 0.510 \\
   \beta_{A_\lambda} &= 548.74 
\end{align}
Since $A_\k=0$, $\beta_{A_\k}$ is not amenable to a numerical extraction.
However we have checked that $\d A_\k$ 
never plays a significant role, we will omit it from our discussion.  The most striking observation is that $\beta_{A_\l}$ is very large, it is practically two orders of magnitude larger than the $\beta$ of all the other parameters. In sharp contrast, note the  tiny $ \beta_{m_\k}$. When we recall that for this point   the ratio $A_t/A_\l$ is very large, $A_t/A_\l \sim 27$, these findings are not surprising, see Eqs.~\ref{approx-beta}. We therefore expect a large scale dependence in the $\drbar$ scheme and probably a large correction for those branching ratios that are most sensitive to $A_\l$. A glance at Fig.~\ref{fig:pointA-paradep} indicates that $h_3^0 \to h_1^0 h_2^0$ and $A_2^0 \to Z h_2^0$ (and its SU(2) equivalent $H^+ \to W^+ h_2^0$) are two such observables.  \\

\noi We have also derived  the  finite parts of the corresponding counterterms. For this benchmark we take  $\bar{\mu} ={\rm Q}_{\rm
susy}=1117.25$ GeV, see Eq.~\ref{eq:finite_ct}. As mentionned in Section ~\ref{sec:benchmark},  to
ensure an {\it a priori} good extraction of the finite parts we computed these finite parts  in the
schemes $t_{134A_1A_2}$ and $OS_{34h_2A_1A_2H+}$. The results are given in Table~\ref{tab:finitePtA}. \\ 
\begin{table}[h]
\noindent\makebox[\linewidth]{\rule{0.8\paperwidth}{0.4pt}}
\caption{\label{tab:finitePtA} Finite parts of the various counterterms which
play a  role in the parametric dependence of the partial widths
computed at $\bar{\mu} ={\rm Q}_{\rm susy}=1117.25$ GeV.}
\begin{center}
 \begin{tabular}{|c|c|c|c|c|c|}
\hline
Scheme & $\d \mu/\mu$ & $\d t_\b/t_\b$ & $\d \l /\l$ & $\d m_\k/m_\k$ & $\d A_\l/A_\l$
\\
\hline 
$t_{134A_1A_2}$&$-2.42\%$ & $0$& $62.26\%$& $-0.67\% $ & $-5.49\% $\\
$OS_{34h_2A_1A_2H+}$ & $-1.57\%$& $-80.69\%$& $-7.88\%$& $0.3\%$& $134\%$\\
\hline
\end{tabular}
\end{center}
\noindent\makebox[\linewidth]{\rule{0.8\paperwidth}{0.4pt}}
\end{table}

\noi We note that at the level of the counterterms, the scheme dependence in $\mu$ and $m_\kappa$ is extremely small (less than one per-cent) and that  in both schemes the values of these two counterterms  are quite small. This is no surprise since $\mu$ can be extracted almost directly from one of the chargino masses while $m_\kappa$ was chosen as an independent parameter precisely because it is an almost direct measure of the singlino mass in this small $\l$ limit. As for the counterterms for $A_\l, \l$ and $\tb$ we have large corrections. In the OS scheme $\d A_\l/A_\l$
is more than $100\%$ and $\d t_\b/t_\b$ is also large. One would think that the use of $M_{H^+}$ as an input in the OS scheme would have constrained $\d A_\l/A_\l$ far better. In fact, as can be derived from the expression of the charged Higgs mass, Eq.~\ref{eq:mh+_tree}, in the limit of small $\l$ and rather large $\tb$, as is the case here,  
$\delta M_{H^+}^2/M_{H^+}^2 \sim (\d A_\l/A_\l+1.8 \d t_\b/t_\b)/2$. Therefore it  is only the combination $(\d A_\l/A_\l+1.8 \d t_\b/t_\b)$ of these two counterterms which is well constrained. This is corroborated  by the values of these two counterterms in Table~\ref{tab:finitePtA}. This issue with  $\tb$  is similar to the one
encountered in the MSSM where the Higgs masses alone are not efficient to
reconstruct $\tb$ \cite{Baro:2008bg}. The good extraction of $\d A_\l/A_\l$ in the $t_{134A_1A_2}$ scheme is therefore a result of  the $\drbar$ condition $\d t_\b/t_\b=0$. This said, the $t_{134A_1A_2}$ scheme gives a bad reconstruction of $\d \l /\l$. As argued in \cite{Belanger:2016tqb}, in this small $\l$ limit, the chargino/neutralino masses are not sensitive to $\lambda$. This leads to a large uncertainty on $\d \l /\l$. The Higgs system with the inclusion of the singlet dominated $h_2^0$ in the OS scheme fares better as demonstrated in Table~\ref{tab:finitePtA}. \\

\noi To summarise, we foresee  {\it i)} in the $\drbar$ scheme, a large scale
variation for decays and couplings that feature a non negligible dependence on
$A_\l$ ($h_3^0 \to h_1^0 h_2^0$, $A_2^0 \to Z h_2^0$ and $H^+ \to W^+ h_2^0$).
The correction should minimise for $\bar \mu =Q_{{\rm susy}}$, {\it ii)} in the
$t_{134A_1A_2}$ scheme the corrections should be mainly driven by $\d \l$. Since
$\beta_\l$ is small, the scale dependence in this mixed scheme is negligible.
All decays are affected expect those not involving any singlet state,  {\it
iii)} in the OS scheme one should pay a special attention to those observables where the
$A_\l$ dependence is important and to a lesser extent the $\tb$ dependence. 

\subsubsection{Full one-loop results}
\begin{table}[!htb]
\noindent\makebox[\linewidth]{\rule{0.8\paperwidth}{0.4pt}}
\begin{center}
\caption{Partial decay widths of Higgs  bosons in other Higgs bosons and/or neutralinos, charginos and gauge bosons at tree-level (in MeV) and the percentage relative full one-loop correction, in the mixed scheme where only $\tb$ is taken $\drbar$ ($t_{134A_1A_2}$) the evaluation is made at the scale $Q_{{\rm susy}}$, the full OS scheme and  full $\drbar$ at two scales, $Q_M$ is taken to be the mass of the decaying particle (see text for details of the schemes).}
\label{tab:higgs:point3}
\begin{tabular}{llllll}\hline 
& {\tiny tree-level}&\multicolumn{4}{c}{One-loop} \\
& (MeV) & $t_{134A_1A_2}$ & $OS_{34h_2A_1A_2H^+}$ & $\drbar\,Q_M$ & $\drbar\,Q_{\rm susy}$\\\hline 
$h_2^0\rightarrow A_1^0A_1^0$ & $47.9$ &  128\% & 
-12\%& 0.4\%& -0.4\%\\\hline
$h_3^0\rightarrow h_1^0h_2^0$ & $22.1$ & 116\% &
79\%& 52\%& -1.7\%\\
$h_3^0\rightarrow \tilde{\chi}_1^0\tilde{\chi}_3^0$ & $35.2$ & 122\% &
-3\% & 2\% & 0.3\% \\
$h_3^0\rightarrow \tilde{\chi}_2^0\tilde{\chi}_3^0$ & $33.8$ & 126\% &
-35\% & 3\% & 1.1\%\\
$h_3^0\rightarrow \tilde{\chi}_1^+\tilde{\chi}_1^-$ & $45.5$ & 1\% &
-11\% & -9\%& -7.4\%\\\hline
$A_2^0\rightarrow Z^0 h_2^0$ & $18.6$ &  120\% &
80\% & -56\% & -14.5\%\\
$A_2^0\rightarrow \tilde{\chi}_1^0\tilde{\chi}_1^0$ & $33.0$ & 28\% &
13\%  & 0.3\% & -1.6\%\\
$A_2^0\rightarrow \tilde{\chi}_1^0\tilde{\chi}_3^0$ & $24.4$ & 130\% &
-31\% & 8\% & 6.2\%\\
$A_2^0\rightarrow \tilde{\chi}_2^0\tilde{\chi}_3^0$ & $30.2$ & 122\%&
-5\% & -0.4\% & -1.9\%\\
$A_2^0\rightarrow \tilde{\chi}_1^+\tilde{\chi}_1^-$ & $55.1$ & -10\% &
-1.5\% & -6\% & -8\%\\\hline
$H^+\rightarrow W^+ h_2^0$ & $20.1$ & 119\%& 79\% &
-56\%& -16\%\\
$H^+\rightarrow \tilde{\chi}_1^+ \tilde{\chi}_3^0$ & $64.0$ & 125\%&
-18\% & 3\% & 1.1\%\\\hline
\end{tabular}
\end{center}
\noindent\makebox[\linewidth]{\rule{0.8\paperwidth}{0.4pt}}
\end{table}
 Full one-loop results for a variety of Higgs decays are shown in Table~\ref{tab:higgs:point3}. For each decay we give the result for the mixed $t_{134A_1A_2}$ scheme, the on-shell scheme $OS_{34h_2A_1A_2H^+}$ and the full $\drbar$ scheme. For the 
 $t_{134A_1A_2}$ scheme  we set the scale at $Q_{\rm susy}=1117.25$GeV. For the
$\drbar$ scheme  we consider both an implementation with a scale at $Q_{\rm
susy}$  as well with a scale $Q_M$ that corresponds to the mass of the decaying
Higgs. For the important decays of $h_3^0, A_2^0,H^+$, $Q_M \sim Q_{\rm
susy}/2$. A quick glance at the table reveals that the  corrections in the
$\drbar$ scheme at $Q_{\rm susy}$ are quite small in practically all channels.
The same results at the scale $Q_M$ are within $2\%$ {\em except} for  the
notable decays $h_3^0\rightarrow h_1^0h_2^0$, $A_2^0 \to Z h_2^0$ and $H^+ \to
W^+ h_2^0$ as a consequence of the very large $\beta_{A_\l}$. The results in the
mixed scheme show a very large and almost common correction of order $120\%$!
(due to the large $\delta \l/\l \sim 62\%$ in this scheme) {\rm except} for
$h_3^0\rightarrow \tilde{\chi}_1^+\tilde{\chi}_1^-$, $A_2^0\rightarrow
\tilde{\chi}_1^+\tilde{\chi}_1^-$ and $A_2^0\rightarrow
\tilde{\chi}_1^0\tilde{\chi}_1^0$. The corrections in the OS scheme are small to
moderate {\em except} for the same notable decays where the scale dependence in
the $\drbar$ scheme is large, $h_3^0\rightarrow h_1^0h_2^0$, $A_2^0 \to Z
h_2^0$ and
$H^+ \to W^+ h_2^0$. Following our discussion on the parametric dependence and
the values of the counterterms, as well as the $\beta$ functions, these results
are easily understood. In fact this is in perfect agreement with the arguments
we summarised at the end of the preceding subsection. \\

\noi The smallness of the radiative corrections in the $\drbar$ scheme seems to
indicate that $\bar \mu \sim Q_{\rm susy}$ is the effective scale for all the
decays in this model. This is also indicative that genuine corrections beyond
the running of the parameters are quite small. We can be more quantitative about
the differences between the schemes and the scale dependencies by combining the
parametric dependencies derived from tree-level considerations in
Eqs.~\ref{varia_h2a1a1_A}-\ref{varia_a2h2Z_A} with the values of the finite
parts of the counterterms in Table.~\ref{tab:finitePtA}, that is $\Delta \ppi
\to \delta \ppi$. Let us go through the results of some of the decays.

\begin{itemize}
\item $h_2^0\rightarrow A_1^0A_1^0$  \\
The scheme  dependence of the relative
correction to the decay is contained in $2 (\delta \l /\l +2 \delta
m_\k/m_\k-\delta \mu/\mu) \sim 2 \delta \lambda/\lambda$. This contribution 
from the finite part of the counterterms gives practically the full one-loop
correction in all the schemes, for instance in the OS scheme this contribution
returns a $-11.4\%$ correction by using the values given in
Table~\ref{tab:finitePtA}. This is another manifestation that genuine
corrections from three-point functions contributions are negligible. The scale
dependence for the $\drbar$ scheme is tiny, indeed with the negligible
$\beta_{m_\k}$ and with $\beta_\l \sim \beta_\mu$ the difference in
the correction between the scale $Q_{\rm susy}$ and $m_{h_2^0}$ (using the
general formula of Eq.~\ref{change-drbar} for the amplitude) is $2
(\beta_\l+2\beta_{m_\k}-\beta_\mu)\ln (m_{h^0_2}^2/Q_{\rm susy}^2) \sim -0.7\%$.
This difference is an excellent approximation to the full one-loop result.

\item $A_2^0\rightarrow \tilde{\chi}_1^0\tilde{\chi}_1^0$,  $A_2^0\rightarrow
\tilde{\chi}_1^+\tilde{\chi}_1^-$ and $h_3^0\rightarrow
\tilde{\chi}_1^+\tilde{\chi}_1^-$ \\ 
We classified these decays in the second category where all particles involved
these decays are MSSM-like though with a very small singlet component for  the
neutralino case. We note that for these decays  the scheme dependence is much
smaller as compared to the other categories. In particular,  these are the only
decays where the relative one-loop corrections are under control in the
$t_{134A_1A_2}$ scheme.  The largest correction in this class shows up for
$A_2^0\rightarrow \tilde{\chi}_1^0\tilde{\chi}_1^0$ in the $t_{134A_1A_2}$
scheme driven essentially by the large value of $\delta \l$ despite the small
parametric dependence of this parameter. Indeed, Eq.~\ref{a2n1n1} when
interpreted in terms of counterterms is an excellent explanation of the results
we find for the full corrections and the scheme dependence. 

\item $h_3^0 \to h_1^0 h_2^0$. \\
Apart from the $\drbar$ scheme
with a scale at $Q_{\rm susy}$ where the correction is modest, all
other schemes  lead to large corrections. We verify, based on the parametric
dependence in Eq.~\ref{varia_h1h2h3_A} and on the fact that $\beta_{A_\l} \gg
\beta_\l \gg \beta_{m_\k}$, that the difference between the scales in $\drbar$
is indeed given 
by $2 \times 0.38 \beta_{A_\l}\ln (m_{h^0_3}^2/Q_{\rm susy}^2) \sim -54\%$. The
values for the corrections in the OS scheme and the mixed scheme are also very
well approximated by the parametric dependence upon replacing the variations by
the corresponding counterterms found in Table.~\ref{tab:finitePtA}. In the OS
scheme the correction is driven essentially by the poor extraction of $A_\l$
while in the mixed scheme it is again essentially the imprecise input $\d \l/\l$
which is behind the large correction.

\item $A_2^0  \to Z^0 h_2^0$ and the equivalent $H^+ \to W^- h_2^0$ \\ 
Because of  the  large running $A_\l$, it is sufficient to consider  $A_\l$ when  comparing the  results in the $\drbar$ scheme at the two
scales. 
The parametric dependence, Eq.~\ref{varia_a2h2Z_A}, 
explains very well  the $\sim 40\%$ difference between the two scales in the $\drbar$
scheme. The difference between the OS scheme and the
$\drbar$ at $Q_{\rm susy}$ is driven essentially by the determination of $A_\l$
and $\tb$ both of which are badly derived in the OS scheme, whereas the
discrepancy in the mixed scheme comes once again from a bad reconstruction of $\l$.

\item All the other remaining  decays of Table~\ref{tab:finitePtA}, 
are those where the Higgses (CP-even, CP-odd or charged) are decaying into
neutralinos/charginos involving the mostly singlet $\neuth$. These decays
would vanish in the $\l \to 0$ limit. We studied the parametric dependence of a
representative of these decays earlier, $h_3^0 \to \neuto \neuth$. We verified
the strong $\l$ dependence, and noted a small $\tb$ dependence (which takes
place in the neutralino/chargino sector). Translated in terms of counterterms,
the parametric dependence is $2 \d \l/\l$ which again explains extremely well
the almost uniform large correction, $\sim 120\%$, in the $t_{134A_1A_2}$
scheme. In the OS scheme, the corrections are much smaller but  the residual
$\tb$ dependence in some of these decays is not totally negligible. 
\end{itemize}

\noi Point A is somehow pathological in the sense that it is MSSM-like and the
amount of mixing is small. This makes it difficult to reconstruct all the
parameters rather precisely. The OS scheme would perform well if it were not 
penalised by a very imprecise reconstruction of $\tb$ which impacts badly on the
reconstruction of $A_\l$. The mass of the charged Higgs as an input only
constrains a very specific combination of these two parameters. The MSSM is
fraught with the same problem of a reconstruction of $\tb$ from the Higgs masses
alone. We have shown that in the MSSM a very good scheme for extracting $\tb$
relied on the decay $A^0 \to \tau^- \tau^-$ \cite{Baro:2008bg}. In the NMSSM,
this issue needs to be investigated in depth and is left for a future work.

\subsection{Point B}
The most crucial features to keep in mind when reviewing the results for Point
B, especially after what we have seen for Point A, is the fact that $\l$ is
large (and $\tb$ small) and $A_\l$ is smaller than $A_t$ by only a factor of 2.
The last observation should mean that $\beta_{A_\l}$ should not be excessively
large. With such value of $\l$ (and $A_\l$) this point constitute a genuine
example of the NMSSM, the branching ratios for the decays we consider here are at least an
order of magnitude larger than in Point A.  The difficulty now is that the
notion of an almost singlet (and MSSM-like) state will be lost, couplings
between the physical states will depend strongly on the pattern of the mixing
matrices. The study of the parametric dependence of the couplings, and hence the
decays, on the underlying parameters is here even more important. This is what
we look at, at tree-level, for a few decays for which we have
calculated the full one-loop corrections. 

\subsubsection{Tree-level. Parameter dependence on some couplings}
\begin{figure}[hbt!]
\begin{center}
\caption{Parameter dependence on (the square) of the couplings that enter some
important decays which we will study at one-loop. We look at the variation in
the parameters $\tb, \l, m_\kappa, A_\kappa, A_\lambda$ and $\mu$. Plotted is
the
percentage variation measured from the reference point. We allow variations of
$\pm 20\%$ for these parameters apart from $A_\kappa$ whose reference value,
$A_\kappa=0$, is varied smoothly up to $40$ GeV. The solid (blue) lines
represents the $h_3^0 h_2^0 h_1^0$
coupling, the long dash-dotted (purple) the $H^+ W^- A_1^0$ 
coupling, the dotted (green) line  the $A_2^0 A_1^0 h_1^0$ coupling, the
dashed (red) line the $h_3^0 h_2^0 h_2^0$ coupling and the dash-dotted
(turquoise) line the $A_2^0 \charg \chargm$ coupling.}
\label{fig:pointB-paradep}
\includegraphics[scale=0.8,trim=0 0 0 0,clip=true]{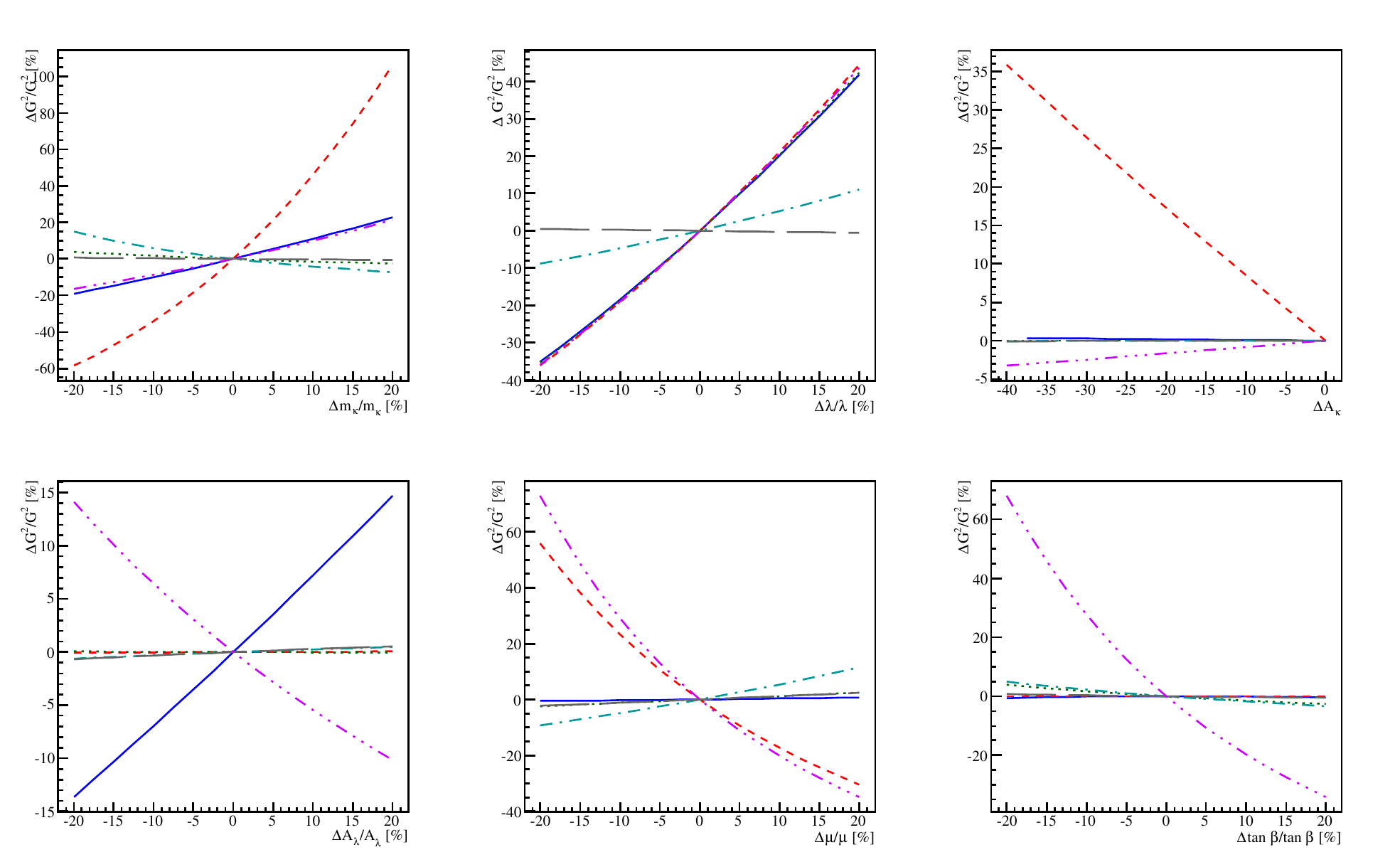}
\end{center}
\end{figure}

\noi Fig.~\ref{fig:pointB-paradep} shows the variations of some couplings when some of the underlying parameters are perturbed within $\pm 20\%$. For Point B we have also examined variations in $M_1$ and $M_2$ which we do not show in Fig.~\ref{fig:pointB-paradep} to avoid clutter. 
The most striking difference with Point A is the fact that for the tri-linear Higgs couplings, $h_3^0 h_1^0 h_2^0$, $h_3^0 h_2^0 h_2^0$ and $h_1^0 A_1^0 A_2^0$, the dependence on almost all parameters is large and highly non linear even for parametric variations of about $10\%$. For example a $10\%$ change in $A_\l$ around its reference value gives a variation of almost $100\%$ on the tri-linear Higgs couplings. Recall here that $\tb$ is small, but we can see that small changes in this
parameter give rise to dramatic changes in the tri-linear coupling. Decays
involving two Higgses and a gauge boson, one such coupling is $A_1^0 H^+ W^-$,
show also large deviations though not as dramatic as for couplings involving
three Higgses. Decays of the Higgs into neutralinos/charginos such as $A_2^0 \to
\chi^+_1 \chi^-_1$ show much more moderate variations with practically an almost
linear dependence. Although not shown in these figures, for decays into
charginos and neutralinos the effect of a change in the gaugino masses $M_1,M_2$
is not totally negligible. For most decays the parameters that lead to the
largest changes are $A_\l,\l,\mu$ and $\tb$.  Considering the highly non-linear
behaviour of these variations, a linear parametrisation is
justified only for small variations. We will therefore first check whether the
finite parts of the counterterms in this model are small enough. If so we will
give the parameterisation as befits a one-loop correction where the counterterms
enter only at first order. \\

\noi In view of these preliminary observations, the 2-body decays studied here
can be classified into 3 categories depending on how many Higgses are taking
part in the decay:

\begin{itemize}
\item a single Higgs for decays into neutralinos and charginos, the amplitude will then  involve the matrix $S_h$.
\item two Higgses and a gauge boson where in this case the Higgses are of opposite parity hence involving the product of the diagonalising matrices $S_h\times P_a$.
\item three Higgses  where   in the case of three neutral CP-even Higgs  the elements of $(S_h)^3$ enter and in the case of a CP-even Higgs decaying into 2 pseudoscalars the amplitudes call for $S_h \times P_a^2$. 
\end{itemize}
The dependence of these  mixing matrices on the underlying parameters, in particular $\l, A_\l$,  
is highly non-linear when $\lambda$ is not small. In the latter case one  expect that   when more and more Higgses are involved, like for instance in the case of the Higgs self-couplings,  the mixing matrices introduce highly non trivial dependencies.

 \subsubsection{Point B: Finite part of the counterterms and their $\beta$ constant}
 We first list  the $\beta$ functions,  in units of $10^{-3}$ they are
\begin{align}
 \beta_{\mu} &= 7.13\\
 \beta_{t_\beta} &= -8.81 \\
 \beta_\lambda &= 10.22\\
\beta_{m_\kappa}& = 6.19\\
 \beta_{A_\lambda} &= 61.36
\end{align}
As expected, since the ratio $A_t/A_\l$ has decreased by about a factor 10
compared to  model A, the value of  $\beta_{A_\l}$  has decreased by almost an
order of magnitude, see Eq.~\ref{approx-beta}. Yet, this is still the largest
$\beta$ constant (6 \% rather than 1\% for the others).

\begin{table}[h]
\noindent\makebox[\linewidth]{\rule{0.8\paperwidth}{0.4pt}}
\caption{\label{tab:finitePtB} Finite parts of the various counterterms which
play a  role in the parametric dependence of the partial widths
computed at $\bar{\mu} ={\rm Q}_{\rm susy}=753.55$ GeV.}
\begin{center}
 \begin{tabular}{|c|c|c|c|c|c|}
\hline
Scheme & $\d \mu/\mu$ & $\d t_\b/t_\b$ & $\d \l /\l$ & $\d m_\k/m_\k$ & $\d A_\l/A_\l$
\\
\hline 
$t_{123A_1A_2}$&$-1.04\%$ & $0$& $3.71\%$& $-1.5\% $ & $6.85\% $\\
$OS_{12h_2A_1A_2H^+}$ & $-1.63\%$& $6.50\%$& $5.94$& $-1.52\%$& $3.40\%$\\
\hline
\end{tabular}
\end{center}
\noindent\makebox[\linewidth]{\rule{0.8\paperwidth}{0.4pt}}
\end{table}

\noi The finite part of the counterterms (evaluated at $Q_{\rm susy} \sim
754$GeV) that we extract are given in Table~\ref{tab:finitePtB}. All
counterterms are of the same order, none exceeding $7\%$ and in any case they
are much smaller than some of  the large values we found for Point A. This
rather precise extraction has to do with the fact that, because of the not so
small mixing, a large number of observables, in particular masses, are quite
sensitive to the underlying parameters. The fact that these values are not very
large justifies parameterising the variation at first order in the counterterm.
The first derivative of the variation, at the origin, gives the infinitesimal
parameterisation. 

\subsubsection{One loop results and analysis of scheme dependence.}
\begin{table}[!htb]
\noindent\makebox[\linewidth]{\rule{0.8\paperwidth}{0.4pt}}
\begin{center}
\caption{Partial decay widths of Higgs  bosons in other Higgs bosons and/or
neutralinos, charginos and gauge bosons at tree-level (in GeV) and the
percentage relative full one-loop correction, in the mixed scheme where only
$\tb$ is taken $\drbar$ ($t_{134A_1A_2}$) at the scale $Q_{{\rm susy}}$, the
full OS scheme, and the full $\drbar$ schemes at two scales. $Q_M$ is
taken to be the mass of the decaying particle.}
\label{tab:higgs:tp4}
\begin{tabular}{llllll}\hline 
& {\tiny tree-level}&\multicolumn{4}{c}{One-loop} \\
& (GeV)  & $t_{123A_1A_2}$ &$OS_{12h_2A_1A_2H^+}$ & $\drbar\,Q_M
$ & $\drbar\,Q_{\rm susy}$\\\hline 
$h_3^0\ra \tilde{\chi}_1^0\tilde{\chi}_2^0$ &0.726 & 13.3\%&
14\%& 5\%&  3\%\\
$h_3^0 \ra A_1^0 Z^0$ &0.613 &  13\%&3\%&  -3\%& 8
\%\\
$h_3^0 \ra h_2^0 h_1^0$ & 0.341&  -142\%&-25\%& 
-106\%&-50\% \\
$h_3^0 \ra h_2^0 h_2^0$ & 0.514 &  51\%&6\% & 13\%& 
-28\%\\
\hline 
$A_2^0 \ra  \tilde{\chi}_1^+\tilde{\chi}_1^-$ & 1.523&  9\%&
7\%&2\% &  1\%\\
$A_2^0 \ra  \tilde{\chi}_1^0\tilde{\chi}_1^0$ &0.723&  19\%&
32\%&  2\%&  2\%\\
$A_2^0 \ra Z^0 h_2^0$ & 0.638 & -10\% &12\%&  -16\%&
-9\% \\
$A_2^0 \ra A_1^0 h_1^0$ & 0.415& -43\%&-0.3\% &  -32\%
&-17\% \\
\hline
$H^+\rightarrow \tilde{\chi}_1^+ \tilde{\chi}_2^0$ & 1.056 & 10\% &
 6\% &  10\% &  8\% \\
$H^+\rightarrow W^+ h_2^0$ & 0.609& -11\% &11\% &  -18\%
&  -10\%\\
$H^+\rightarrow W^+ A_1^0$ & 0.603 & 12\% &2\%&  -3\%&
 -9\%\\
$H^+\rightarrow \tilde{\chi}_1^+  \tilde{\chi}_1^0$ & 0.561&  14\% &
21\%& 9\%&  9\%\\
\hline
\end{tabular}
\end{center}
\noindent\makebox[\linewidth]{\rule{0.8\paperwidth}{0.4pt}}
\end{table}

\noi A quick inspection of the results in Table~\ref{tab:higgs:tp4} reveals
that, if we leave out the decays of the category involving the tri-linear  Higgs
couplings, in particular the CP-even $h_3^0 \to h_1^0 h_2^0, h_2^0 h_2^0$, the
radiative corrections are moderate, especially compared to Point A. Overall,
the OS scheme performs quite well. In particular, the OS scheme  returns the
smallest corrections for the problematic decays $h_3^0 \to h_1^0 h_2^0, h_2^0
h_2^0$. Still, in most cases the scheme dependence is not negligible. This is
not surprising considering the abrupt variations we observed at tree-level on
the Higgs trilinear couplings.

\begin{itemize}
\item Decays into charginos and neutralinos ($A^0_2 \to \charg \chargm$,
$A_2^0 \to \neuto \neuto$) \\
 First observe that the scale dependence in these decays is never larger than
$2\%$. This is due in a large part to the fact that the couplings involved in
these decays are insensitive to $A_\l$, a parameter which comes with the largest
$\beta$. The $\beta$ constants for the other parameters are all smaller than
$1\%$. Besides, as 
Fig.~\ref{fig:pointB-paradep} confirms for $A^0_2 \to \charg \chargm$, the 
parametric dependence is quite small for these decays
compared to those where more than one Higgs is involved. We can write 
\beqn
\frac{ \delta \Gamma_{A_2^0 \to \charg \chargm}}{\Gamma_{A_2^0 \to \chi^+_1
\chi^-_1}^0}&\sim& 0.67\frac{\delta\l}{\l} +0.55 \frac{\delta\mu}{\mu}
-0.3 \frac{\delta{t_\b}}{t_\b} 
\end{eqnarray}
The differences between schemes are within about $10\%$, the smallest
corrections are usually
obtained in the $\drbar$ scheme.  The scheme dependence is quite small for all the decays in this category  apart from the special case of $A_2^0 \to \neuto \neuto$. $\neuto$ here is  bino-dominated but with a  large wino and higgsino component. 
In this case we have worked out the parametric dependence including the $M_1$ and $M_2$ counterterms
\begin{eqnarray}
\frac{ \delta \Gamma_{A_2^0 \to \neuto \neuto}}{\Gamma_{A_2^0 \to \neuto
\neuto}^0}&\sim& 1.1\frac{\delta\l}{\l} -2.4 \frac{\delta\mu}{\mu}
+0.9 \frac{\delta{t_\b}}{t_\b}  +2.6 \frac{\delta M_1}{M_1} - 0.9\frac{\delta
M_2}{M_2}-0.9 \frac{\d m_\k}{m_\k}
\end{eqnarray}
The $\sim 10\%$ difference between the $t_{123A_1A_2}$ and the OS scheme is
essentially due to the $\tb$ definition in the two schemes.  The relatively
large correction of $32\%$ in the OS scheme is in fact due to the addition of
many smaller contributions  including $M_1$, $M_2$ (and $\mu$) which all affect
the neutralino sector. 

\item Higgs decays into  a vector boson  and another Higgs ($A_2 \to Z^0 h_2^0,
H^+ \to W^+ h_2^0, H^+ \to W^+A_1^0,h_3^0 \to A_1^0 Z^0$ )\\
The pattern of the corrections for these decays is quite similar. In all the
schemes, the one-loop corrections are moderate. They are, in absolute terms, 
within $20\%$ with a scheme difference that can attain $30\%$. In $\drbar$ the
scale dependence is about $6\%$ when we compare the values obtained  at $Q_{\rm
susy} \sim 754$ GeV and at a scale about the mass of
the decaying Higgs $\sim 480$ GeV. For $H^+ \to W^+ A_1^0$, this is accounted for 
by the $A_\l$ running $\kappa_{A_\l}^{HWA} \beta_{A_\l} \ln (Q_{\rm
SUSY}^2/M_{H^{\pm}}^2)$ with $ \kappa_{A_\l}^{HWA} \sim 1$ as can be inferred
from Fig.~\ref{fig:pointB-paradep}. 

\item $h_3^0 \rightarrow h_2^0 h_1^0$\\
 With three Higgses involved in the
process, the parametric dependence becomes very important. Because of the large
value of $\beta_{A_\l}$ compared to all other $\beta$'s, one still expects the
running to be dominated by this parameter. Indeed, the corrections in the $\drbar$
schemes are not only quite large, $-50\%$ at $Q_{\rm susy}\sim 754$GeV, but they
are also very
sensitive to the choice of scale since at the scale $M_{H^+}$ which is not even
half $Q_{\rm susy}$ the corrections more than doubles to $-106\%$. This is 
driven by   the large variation due to  $A_\l$ as
observed in Fig.~\ref{fig:pointB-paradep}. 
One can approximate the variation as $10 \beta_{A_\l}  \ln (Q_{\rm
SUSY}^2/M_{h^0_3}^{2})\sim 60\%$. In fact the  dependence in the counterterms
can be worked out more precisely 
\begin{eqnarray}
\frac{ \delta \Gamma_{h_3^0 h_2^0 h_1^0}}{\Gamma_{h_3^0 h_2^0
h_1^0}^0}&\sim& 1.3 \frac{\delta\l}{\l}+ 16
\frac{\delta\mu}{\mu}+11\left( 1.15 \frac{\delta{t_\b}}{t_\b}
-\frac{\delta{A_\l}}{A_\l}\right) +0.2 \frac{\d
m_\k}{m_\k}
\end{eqnarray}

With the extracted values of the counterterms in the OS and mixed schemes
(written
for  $\bar{\mu}=Q_{\rm susy}$) we  recover the differences shown in
Table~\ref{tab:higgs:tp4} between the OS and $\drbar$ and $t$ schemes.  Note
also the ``compensation" between the  variation in $\tb$ and $A_\l$ in the
parametric dependence which is effective in keeping the correction in the OS
scheme manageable. 

Although the scheme dependence is well understood, the fact that the corrections
in the $\drbar$ scheme at the scale $Q_{{\rm susy}}$ are large indicates that
this scale is not the most appropriate effective scale which minimises the
correction. In this respect the OS scheme gives a more ``perturbative"
prediction. Nonetheless a scale of $Q_{{\rm eff.}} \sim 1.5 Q_{{\rm susy}}$
gives a correction of about only $1\%$. An effective scale  $\sim 1.2 Q_{{\rm
susy}}$ reproduces the result of the OS scheme. This shows that small changes in
the scale (around $Q_{{\rm susy}}$ ) reduce the results quite significantly.
This is driven mostly by the large sensitivity in some of the parameters,
essentially $A_\l$ whose $\beta$ constant is the largest. 

\item $h_3^0 \to h_2^0 h_2^0$ \\ 
The parametric dependence that tracks the
dominant variations are also very large here as we remarked earlier. It can be approximated as 

\begin{eqnarray}
\label{param_h3h2h2}
\frac{ \delta \Gamma_{h_3^0 h_2^0 h_2^0}}{\Gamma_{h_3^0 h_2^0
h_2^0}^0}&\sim&
4 \frac{\delta\l}{\l} -8 \left(\frac{\delta\mu}{\mu}
+\frac{1}{2}\frac{\delta{t_\b}}{t_\b}
-\frac{\delta{A_\l}}{A_\l}\right) 
+0.3\frac{\d m_\k}{m_\k} 
\end{eqnarray}
We note that although the $A_\l$ dependence is large it is not as large as the
one found for the $h_1^0 h_2^0 h_3^0$ coupling. Indeed, before performing the
diagonalisation to the physical basis, this coupling would stem from the $h_s^0
h_s^0 h_d^0$ part of the potential whose strength is $\l^2 v_d \propto \l^2
c_\beta$, This also explains the quartic dependence on $\l$ of the coupling. 
Even though the  parametric dependence on $A_\l$  is more
moderate than for $h_3^0 h_2^0 h_1^0$, it remains a strong parametric
dependence in this decay also. Add to this the fact that $\beta_{A_\l}$ is the
largest of all $\beta$, the large scale dependence is driven essentially by
$\beta_{A_\l}$.
For the coupling responsible for this decay,  an effective scale about
twice lower than what was found in  $h_3^0 h_2^0 h_1^0$ is required to bring the
corrections to a negligible level. Varying again by $30\%$ $Q_{{\rm susy}}$
brings the $\drbar$ corrections in par with the correction in the OS scheme. 
Eq.~\ref{param_h3h2h2} when combined with the values of the finite part of the
counterterms in the OS scheme given in Table~\ref{tab:finitePtB} reproduce very
well the difference between the OS scheme and the $\drbar$ scheme. 

\item $A_2^0 \ra A_1^0 h_1^0$\\
For this decay the parametric dependence can be approximated by
\begin{eqnarray}
\frac{ \delta \Gamma_{A_2^0 \to A_1^0 h_1^0}}{\Gamma_{A_2^0 \to A_1^0
h_1^0}}&\sim&0.9 \frac{\delta\l}{\l} + 8.4
\left(\frac{\delta\mu}{\mu}+\frac{9}{16}\frac{\delta{t_\b}}{t_\b}-\frac{7}{16}
\frac{\delta{A_\l}}{A_\l}\right) -1.1\frac{\d m_\k}{m_\k}.
\end{eqnarray}
Once again, the correction for this tri-linear Higgs coupling is smallest in the
OS scheme. The scale dependence that can be seen from the two values in
the $\drbar$ scheme is not small. As has been a pattern for other tri-linear
Higgs couplings, the rather large scale dependence is also a sign of a large 
correction in the mixed $t_{123A_1A_2}$ scheme. The scale dependence is driven
essentially by $\beta_{A_\l}$. $Q_{{\rm eff.}} \sim 1.5 Q_{{\rm susy}}$ is the
effective scale where the corrections vanishes in the $\drbar$ scheme. Note that
this is the same effective scale we found for $h_3^0 \to h_1^0 h_1^0$. 
The small correction in the OS scheme is a result of a cancellation between the contribution of the $A_\l$ and $\tb$ counterterms. Because in the $t$-scheme, $\tb$ is defined in $\drbar$, this cancellation is not operative and again the correction is dominated by $A_\l$. 
\end{itemize}

\noi To summarise the results for model B, it is worth stressing that in decays
of Higgs into Higgses the OS scheme performs quite well in the sense that it
gives  very small corrections. There is a large scale dependence for these
decays, but the effective scale where the correction vanishes in $\drbar$ is,
after all, not that much different from $Q_{{\rm susy}}$. Although
$\beta_{A_\l}$ is about $6\%$, a value 10 times smaller than in Model A, the
parametric dependence in Model B is strong thus enhancing the loop correction.
Decays into charginos and neutralinos being much less sensitive to $A_\l$ do not
show much scheme and scale dependence.

\section{Conclusions}
\label{sec:conclusions}
With the renormalisation of the Higgs sector, it is now possible to
compute full one-loop electroweak and QCD corrections to masses, decays
and scattering processes in the NMSSM with {\tt SloopS}. This is particularly
relevant as experiments are improving the precision in the measurements of Higgs and Dark Matter
observables. Our computation of partial decay widths illustrates the importance
of pure electroweak corrections for Higgs decays into supersymmetric particles and highlights the choice of 
the renormalisation scheme. Our setup allows to choose between the $\drbar$
scheme, different on-shell schemes and ``mixed" schemes whereby some conditions
are imposed on on-shell quantities and others taken as $\drbar$. 
This variety of schemes has   been implemented within {\tt SloopS}. Comparing
different renormalisation schemes is crucial to weigh the theoretical
uncertainties and the possible necessity of higher order corrections. For this
purpose, we discussed at length how the choice of the minimal set of physical
masses to reconstruct the underlying parameters can affect the numerical results
and their reliability. We have found large radiative corrections for some
observables. Some of these large corrections appear only in certain
renormalisation schemes. In this case, when the scheme dependence is large this
can also be accompanied by a large scale dependence in the $\drbar$ scheme. 
These large scheme dependencies and large scale variations are due to a large
value of some  $\beta$ constants for some specific underlying parameters and/or
are associated with a large parametric dependence of the observable upon this
specific parameter. The latter situation occurs in a NMSSM with a moderate $\l$.
In the small $\l$ scenarios, this parametric dependence is not as large, however
many counterterms are poorly reconstructed precisely because they are extracted
from a set of input masses with little sensitivity on some of the underlying
parameters. It has to be stressed  that,  although easier to implement, taking
only masses as inputs may not be
the optimal choice to renormalise the model. When new particles are discovered,
not only their masses will be measured but so will  the strength of their
production modes and their decays. These observables will thus offer new
possibilities for reconstructing the fundamental parameters of the model that
will  not require the knowledge of the complete particle spectrum. 
It remains to be seen  whether  a more cleverly chosen renormalisation scheme,
for example one that uses the partial width of a heavy Higgs decay as a
renormalisation condition, would lead to better controlled corrections.  In the MSSM we have shown\cite{Baro:2008bg} that the decay of the pseudoscalar Higgs to 
a pair of $\tau$'s is an excellent definition of $\tb$. Some of the large
corrections we found are also pathological in the sense that they are due to a
rather large value of $A_t$ compared to the NMSSM parameter $A_\lambda$. Such a
discrepancy between $A_t$ and $A_\l$ is responsible for a large $\beta_{A_\l}$
which will then propagate into  the corrections of many Higgs observables, in
particular those for the Higgs self-couplings. A natural NMSSM should not
require very large values of $A_t$ as what is required for MSSM-like models
(with $\l \ll 1$), in this case the one-loop corrections are contained and an
on-shell scheme is a quite judicious choice.  

\section*{Acknowledgements}
This research was supported in part by the Research Executive Agency (REA) of
the European Union under the Grant Agreement PITN-GA2012-316704 ("HiggsTools"), by
``Investissements  d'avenir,  Labex   ENIGMASS"  and  by  the  French   ANR,   Project
DMAstro-LHC,   ANR-12-BS05-006.
G.C is supported in part by the ERC advanced grant Higgs@LHC and the
THEORY-LHC France initiative of CNRS/IN2P3. The authors would like to thank A.
Djouadi for useful discussions regarding the project.


\begin{thebibliography}{10}

\bibitem{Aad:2012tfa}
{\bfseries ATLAS} Collaboration, G.~Aad {\em et~al.}, ``{Observation of a new
  particle in the search for the Standard Model Higgs boson with the ATLAS
  detector at the LHC},''
  \href{http://dx.doi.org/10.1016/j.physletb.2012.08.020}{{\em Phys. Lett.}
  {\bfseries B716} (2012) 1--29},
\href{http://arxiv.org/abs/1207.7214}{{\ttfamily arXiv:1207.7214 [hep-ex]}}.

\bibitem{Chatrchyan:2012xdj}
{\bfseries CMS} Collaboration, S.~Chatrchyan {\em et~al.}, ``{Observation of a
  new boson at a mass of 125 GeV with the CMS experiment at the LHC},''
  \href{http://dx.doi.org/10.1016/j.physletb.2012.08.021}{{\em Phys. Lett.}
  {\bfseries B716} (2012) 30--61},
\href{http://arxiv.org/abs/1207.7235}{{\ttfamily arXiv:1207.7235 [hep-ex]}}.

\bibitem{Hall:2011aa}
L.~J. Hall, D.~Pinner, and J.~T. Ruderman, ``{A Natural SUSY Higgs Near 126
  GeV},'' \href{http://dx.doi.org/10.1007/JHEP04(2012)131}{{\em JHEP}
  {\bfseries 04} (2012) 131},
\href{http://arxiv.org/abs/1112.2703}{{\ttfamily arXiv:1112.2703 [hep-ph]}}.

\bibitem{Ellwanger:2009dp}
U.~Ellwanger, C.~Hugonie, and A.~M. Teixeira, ``{The Next-to-Minimal
  Supersymmetric Standard Model},''
  \href{http://dx.doi.org/10.1016/j.physrep.2010.07.001}{{\em Phys. Rept.}
  {\bfseries 496} (2010) 1--77},
\href{http://arxiv.org/abs/0910.1785}{{\ttfamily arXiv:0910.1785 [hep-ph]}}.

\bibitem{Ellwanger:2006rm}
U.~Ellwanger and C.~Hugonie, ``{The Upper bound on the lightest Higgs mass in
  the NMSSM revisited},''
  \href{http://dx.doi.org/10.1142/S0217732307023870}{{\em Mod. Phys. Lett.}
  {\bfseries A22} (2007) 1581--1590},
\href{http://arxiv.org/abs/hep-ph/0612133}{{\ttfamily arXiv:hep-ph/0612133
  [hep-ph]}}.

\bibitem{Ellwanger:2011mu}
U.~Ellwanger, G.~Espitalier-Noel, and C.~Hugonie, ``{Naturalness and Fine
  Tuning in the NMSSM: Implications of Early LHC Results},''
  \href{http://dx.doi.org/10.1007/JHEP09(2011)105}{{\em JHEP} {\bfseries 09}
  (2011) 105},
\href{http://arxiv.org/abs/1107.2472}{{\ttfamily arXiv:1107.2472 [hep-ph]}}.

\bibitem{BasteroGil:2000bw}
M.~Bastero-Gil, C.~Hugonie, S.~F. King, D.~P. Roy, and S.~Vempati, ``{Does LEP
  prefer the NMSSM?},''
  \href{http://dx.doi.org/10.1016/S0370-2693(00)00930-8}{{\em Phys. Lett.}
  {\bfseries B489} (2000) 359--366},
\href{http://arxiv.org/abs/hep-ph/0006198}{{\ttfamily arXiv:hep-ph/0006198
  [hep-ph]}}.

\bibitem{Ross:2012nr}
G.~G. Ross, K.~Schmidt-Hoberg, and F.~Staub, ``{The Generalised NMSSM at One
  Loop: Fine Tuning and Phenomenology},''
  \href{http://dx.doi.org/10.1007/JHEP08(2012)074}{{\em JHEP} {\bfseries 08}
  (2012) 074},
\href{http://arxiv.org/abs/1205.1509}{{\ttfamily arXiv:1205.1509 [hep-ph]}}.

\bibitem{Dittmaier:2011ti}
{\bfseries LHC Higgs Cross Section Working Group} Collaboration, S.~Dittmaier
  {\em et~al.}, ``{Handbook of LHC Higgs Cross Sections: 1. Inclusive
  Observables},''
\href{http://arxiv.org/abs/1101.0593}{{\ttfamily arXiv:1101.0593 [hep-ph]}}.

\bibitem{Dittmaier:2012vm}
S.~Dittmaier {\em et~al.}, ``{Handbook of LHC Higgs Cross Sections: 2.
  Differential Distributions},''
\href{http://arxiv.org/abs/1201.3084}{{\ttfamily arXiv:1201.3084 [hep-ph]}}.

\bibitem{Heinemeyer:2013tqa}
{\bfseries LHC Higgs Cross Section Working Group} Collaboration, J.~R. Andersen
  {\em et~al.}, ``{Handbook of LHC Higgs Cross Sections: 3. Higgs
  Properties},''
\href{http://arxiv.org/abs/1307.1347}{{\ttfamily arXiv:1307.1347 [hep-ph]}}.

\bibitem{deFlorian:2016spz}
{\bfseries LHC Higgs Cross Section Working Group} Collaboration, D.~de~Florian
  {\em et~al.}, ``{Handbook of LHC Higgs Cross Sections: 4. Deciphering the
  Nature of the Higgs Sector},''
\href{http://arxiv.org/abs/1610.07922}{{\ttfamily arXiv:1610.07922 [hep-ph]}}.

\bibitem{Haber:1990aw}
H.~E. Haber and R.~Hempfling, ``{Can the mass of the lightest Higgs boson of
  the minimal supersymmetric model be larger than m(Z)?},''
\href{http://dx.doi.org/10.1103/PhysRevLett.66.1815}{{\em Phys. Rev. Lett.}
  {\bfseries 66} (1991) 1815--1818}.

\bibitem{Ellis:1990nz}
J.~R. Ellis, G.~Ridolfi, and F.~Zwirner, ``{Radiative corrections to the masses
  of supersymmetric Higgs bosons},''
\href{http://dx.doi.org/10.1016/0370-2693(91)90863-L}{{\em Phys. Lett.}
  {\bfseries B257} (1991) 83--91}.

\bibitem{Djouadi:2005gj}
A.~Djouadi, ``{The Anatomy of electro-weak symmetry breaking. II. The Higgs
  bosons in the minimal supersymmetric model},''
  \href{http://dx.doi.org/10.1016/j.physrep.2007.10.005}{{\em Phys. Rept.}
  {\bfseries 459} (2008) 1--241},
\href{http://arxiv.org/abs/hep-ph/0503173}{{\ttfamily arXiv:hep-ph/0503173
  [hep-ph]}}.

\bibitem{Carena:2002es}
M.~Carena and H.~E. Haber, ``{Higgs boson theory and phenomenology},''
  \href{http://dx.doi.org/10.1016/S0146-6410(02)00177-1}{{\em Prog. Part. Nucl.
  Phys.} {\bfseries 50} (2003) 63--152},
\href{http://arxiv.org/abs/hep-ph/0208209}{{\ttfamily arXiv:hep-ph/0208209
  [hep-ph]}}.

\bibitem{Martin:2004kr}
S.~P. Martin, ``{Strong and Yukawa two-loop contributions to Higgs scalar boson
  self-energies and pole masses in supersymmetry},''
  \href{http://dx.doi.org/10.1103/PhysRevD.71.016012}{{\em Phys. Rev.}
  {\bfseries D71} (2005) 016012},
\href{http://arxiv.org/abs/hep-ph/0405022}{{\ttfamily arXiv:hep-ph/0405022
  [hep-ph]}}.

\bibitem{Heinemeyer:2004ms}
S.~Heinemeyer, ``{MSSM Higgs physics at higher orders},''
  \href{http://dx.doi.org/10.1142/S0217751X06031028}{{\em Int. J. Mod. Phys.}
  {\bfseries A21} (2006) 2659--2772},
\href{http://arxiv.org/abs/hep-ph/0407244}{{\ttfamily arXiv:hep-ph/0407244
  [hep-ph]}}.

\bibitem{Ellwanger:1993hn}
U.~Ellwanger, ``{Radiative corrections to the neutral Higgs spectrum in
  supersymmetry with a gauge singlet},''
  \href{http://dx.doi.org/10.1016/0370-2693(93)91431-L}{{\em Phys. Lett.}
  {\bfseries B303} (1993) 271--276},
\href{http://arxiv.org/abs/hep-ph/9302224}{{\ttfamily arXiv:hep-ph/9302224
  [hep-ph]}}.

\bibitem{Elliott:1993uc}
T.~Elliott, S.~F. King, and P.~L. White, ``{Squark contributions to Higgs boson
  masses in the next-to-minimal supersymmetric standard model},''
  \href{http://dx.doi.org/10.1016/0370-2693(93)91321-D}{{\em Phys. Lett.}
  {\bfseries B314} (1993) 56--63},
\href{http://arxiv.org/abs/hep-ph/9305282}{{\ttfamily arXiv:hep-ph/9305282
  [hep-ph]}}.

\bibitem{Elliott:1993bs}
T.~Elliott, S.~F. King, and P.~L. White, ``{Radiative corrections to Higgs
  boson masses in the next-to-minimal supersymmetric Standard Model},''
  \href{http://dx.doi.org/10.1103/PhysRevD.49.2435}{{\em Phys. Rev.} {\bfseries
  D49} (1994) 2435--2456},
\href{http://arxiv.org/abs/hep-ph/9308309}{{\ttfamily arXiv:hep-ph/9308309
  [hep-ph]}}.

\bibitem{Pandita:1993hx}
P.~N. Pandita, ``{One loop radiative corrections to the lightest Higgs scalar
  mass in nonminimal supersymmetric Standard Model},''
\href{http://dx.doi.org/10.1016/0370-2693(93)90137-7}{{\em Phys. Lett.}
  {\bfseries B318} (1993) 338--346}.

\bibitem{Pandita:1993tg}
P.~N. Pandita, ``{Radiative corrections to the scalar Higgs masses in a
  nonminimal supersymmetric Standard Model},''
\href{http://dx.doi.org/10.1007/BF01562550}{{\em Z. Phys.} {\bfseries C59}
  (1993) 575--584}.

\bibitem{Ellwanger:2005fh}
U.~Ellwanger and C.~Hugonie, ``{Yukawa induced radiative corrections to the
  lightest Higgs boson mass in the NMSSM},''
  \href{http://dx.doi.org/10.1016/j.physletb.2005.07.039}{{\em Phys. Lett.}
  {\bfseries B623} (2005) 93--103},
\href{http://arxiv.org/abs/hep-ph/0504269}{{\ttfamily arXiv:hep-ph/0504269
  [hep-ph]}}.

\bibitem{Ender:2011qh}
K.~Ender, T.~Graf, M.~Muhlleitner, and H.~Rzehak, ``{Analysis of the NMSSM
  Higgs Boson Masses at One-Loop Level},''
  \href{http://dx.doi.org/10.1103/PhysRevD.85.075024}{{\em Phys.Rev.}
  {\bfseries D85} (2012) 075024},
\href{http://arxiv.org/abs/1111.4952}{{\ttfamily arXiv:1111.4952 [hep-ph]}}.

\bibitem{Degrassi:2009yq}
G.~Degrassi and P.~Slavich, ``{On the radiative corrections to the neutral
  Higgs boson masses in the NMSSM},''
  \href{http://dx.doi.org/10.1016/j.nuclphysb.2009.09.018}{{\em Nucl. Phys.}
  {\bfseries B825} (2010) 119--150},
\href{http://arxiv.org/abs/0907.4682}{{\ttfamily arXiv:0907.4682 [hep-ph]}}.

\bibitem{Goodsell:2014pla}
M.~D. Goodsell, K.~Nickel, and F.~Staub, ``{Two-loop corrections to the Higgs
  masses in the NMSSM},''
  \href{http://dx.doi.org/10.1103/PhysRevD.91.035021}{{\em Phys. Rev.}
  {\bfseries D91} (2015) 035021},
\href{http://arxiv.org/abs/1411.4665}{{\ttfamily arXiv:1411.4665 [hep-ph]}}.

\bibitem{Staub:2015aea}
F.~Staub, P.~Athron, U.~Ellwanger, R.~Grober, M.~Muhlleitner, P.~Slavich, and
  A.~Voigt, ``{Higgs mass predictions of public NMSSM spectrum generators},''
\href{http://arxiv.org/abs/1507.05093}{{\ttfamily arXiv:1507.05093 [hep-ph]}}.

\bibitem{Drechsel:2016jdg}
P.~Drechsel, L.~Galeta, S.~Heinemeyer, and G.~Weiglein, ``{Precise Predictions
  for the Higgs-Boson Masses in the NMSSM},''
  \href{http://dx.doi.org/10.1140/epjc/s10052-017-4595-1}{{\em Eur. Phys. J.}
  {\bfseries C77} no.~1, (2017) 42},
\href{http://arxiv.org/abs/1601.08100}{{\ttfamily arXiv:1601.08100 [hep-ph]}}.

\bibitem{Drechsel:2016htw}
P.~Drechsel, R.~Groeber, S.~Heinemeyer, M.~M. Muhlleitner, H.~Rzehak, and
  G.~Weiglein, ``{Higgs-Boson Masses and Mixing Matrices in the NMSSM: Analysis
  of On-Shell Calculations},''
\href{http://arxiv.org/abs/1612.07681}{{\ttfamily arXiv:1612.07681 [hep-ph]}}.

\bibitem{Liebler:2015bka}
S.~Liebler, ``{Neutral Higgs production at proton colliders in the
  CP-conserving NMSSM},''
  \href{http://dx.doi.org/10.1140/epjc/s10052-015-3432-7}{{\em Eur. Phys. J.}
  {\bfseries C75} no.~5, (2015) 210},
\href{http://arxiv.org/abs/1502.07972}{{\ttfamily arXiv:1502.07972 [hep-ph]}}.

\bibitem{Baglio:2013vya}
J.~Baglio, T.~N. Dao, R.~Gr{\"o}ber, M.~M. M{\"u}hlleitner, H.~Rzehak,
  M.~Spira, J.~Streicher, and K.~Walz, ``{A new implementation of the NMSSM
  Higgs boson decays},''
\href{http://dx.doi.org/10.1051/epjconf/20134912001}{{\em EPJ Web Conf.}
  {\bfseries 49} (2013) 12001}.

\bibitem{Baglio:2015noa}
J.~Baglio, C.~O. Krauss, M.~M{\"u}hlleitner, and K.~Walz, ``{Next-to-Leading
  Order NMSSM Decays with CP-odd Higgs Bosons and Stops},''
\href{http://arxiv.org/abs/1505.07125}{{\ttfamily arXiv:1505.07125 [hep-ph]}}.

\bibitem{Nhung:2013lpa}
D.~T. Nhung, M.~Muhlleitner, J.~Streicher, and K.~Walz, ``{Higher Order
  Corrections to the Trilinear Higgs Self-Couplings in the Real NMSSM},''
  \href{http://dx.doi.org/10.1007/JHEP11(2013)181}{{\em JHEP} {\bfseries 11}
  (2013) 181},
\href{http://arxiv.org/abs/1306.3926}{{\ttfamily arXiv:1306.3926 [hep-ph]}}.

\bibitem{Baglio:2013iia}
J.~Baglio, R.~Gr{\"o}ber, M.~M{\"u}hlleitner, D.~T. Nhung, H.~Rzehak, M.~Spira,
  J.~Streicher, and K.~Walz, ``{NMSSMCALC: A Program Package for the
  Calculation of Loop-Corrected Higgs Boson Masses and Decay Widths in the
  (Complex) NMSSM},'' \href{http://dx.doi.org/10.1016/j.cpc.2014.08.005}{{\em
  Comput. Phys. Commun.} {\bfseries 185} no.~12, (2014) 3372--3391},
\href{http://arxiv.org/abs/1312.4788}{{\ttfamily arXiv:1312.4788 [hep-ph]}}.

\bibitem{Ellwanger:2005dv}
U.~Ellwanger and C.~Hugonie, ``{NMHDECAY 2.0: An Updated program for sparticle
  masses, Higgs masses, couplings and decay widths in the NMSSM},''
  \href{http://dx.doi.org/10.1016/j.cpc.2006.04.004}{{\em Comput. Phys.
  Commun.} {\bfseries 175} (2006) 290--303},
\href{http://arxiv.org/abs/hep-ph/0508022}{{\ttfamily arXiv:hep-ph/0508022
  [hep-ph]}}.

\bibitem{Ellwanger:2006rn}
U.~Ellwanger and C.~Hugonie, ``{NMSPEC: A Fortran code for the sparticle and
  Higgs masses in the NMSSM with GUT scale boundary conditions},''
  \href{http://dx.doi.org/10.1016/j.cpc.2007.05.001}{{\em Comput. Phys.
  Commun.} {\bfseries 177} (2007) 399--407},
\href{http://arxiv.org/abs/hep-ph/0612134}{{\ttfamily arXiv:hep-ph/0612134
  [hep-ph]}}.

\bibitem{Staub:2010ty}
F.~Staub, W.~Porod, and B.~Herrmann, ``{The Electroweak sector of the NMSSM at
  the one-loop level},'' \href{http://dx.doi.org/10.1007/JHEP10(2010)040}{{\em
  JHEP} {\bfseries 10} (2010) 040},
\href{http://arxiv.org/abs/1007.4049}{{\ttfamily arXiv:1007.4049 [hep-ph]}}.

\bibitem{Porod:2011nf}
W.~Porod and F.~Staub, ``{SPheno 3.1: Extensions including flavour, CP-phases
  and models beyond the MSSM},''
  \href{http://dx.doi.org/10.1016/j.cpc.2012.05.021}{{\em Comput. Phys.
  Commun.} {\bfseries 183} (2012) 2458--2469},
\href{http://arxiv.org/abs/1104.1573}{{\ttfamily arXiv:1104.1573 [hep-ph]}}.

\bibitem{Das:2011dg}
D.~Das, U.~Ellwanger, and A.~M. Teixeira, ``{NMSDECAY: A Fortran Code for
  Supersymmetric Particle Decays in the Next-to-Minimal Supersymmetric Standard
  Model},'' \href{http://dx.doi.org/10.1016/j.cpc.2011.11.021}{{\em Comput.
  Phys. Commun.} {\bfseries 183} (2012) 774--779},
\href{http://arxiv.org/abs/1106.5633}{{\ttfamily arXiv:1106.5633 [hep-ph]}}.

\bibitem{Allanach:2013kza}
B.~C. Allanach, P.~Athron, L.~C. Tunstall, A.~Voigt, and A.~G. Williams,
  ``{Next-to-Minimal SOFTSUSY},''
  \href{http://dx.doi.org/10.1016/j.cpc.2014.04.015}{{\em Comput. Phys.
  Commun.} {\bfseries 185} (2014) 2322--2339},
\href{http://arxiv.org/abs/1311.7659}{{\ttfamily arXiv:1311.7659 [hep-ph]}}.

\bibitem{Athron:2014yba}
P.~Athron, J.-h. Park, D.~St{\"o}ckinger, and A.~Voigt, ``{FlexibleSUSY - A
  spectrum generator generator for supersymmetric models},''
  \href{http://dx.doi.org/10.1016/j.cpc.2014.12.020}{{\em Comput. Phys.
  Commun.} {\bfseries 190} (2015) 139--172},
\href{http://arxiv.org/abs/1406.2319}{{\ttfamily arXiv:1406.2319 [hep-ph]}}.

\bibitem{Goodsell:2017pdq}
M.~D. Goodsell, S.~Liebler, and F.~Staub, ``{Generic calculation of two-body
  partial decay widths at the full one-loop level},''
\href{http://arxiv.org/abs/1703.09237}{{\ttfamily arXiv:1703.09237 [hep-ph]}}.

\bibitem{Belanger:2016tqb}
G.~Belanger, V.~Bizouard, F.~Boudjema, and G.~Chalons, ``{One-loop
  renormalization of the NMSSM in SloopS: The neutralino-chargino and sfermion
  sectors},'' \href{http://dx.doi.org/10.1103/PhysRevD.93.115031}{{\em Phys.
  Rev.} {\bfseries D93} no.~11, (2016) 115031},
\href{http://arxiv.org/abs/1602.05495}{{\ttfamily arXiv:1602.05495 [hep-ph]}}.

\bibitem{Baro:2008bg}
N.~Baro, F.~Boudjema, and A.~Semenov, ``{Automatised full one-loop
  renormalisation of the MSSM. I. The Higgs sector, the issue of tan(beta) and
  gauge invariance},'' \href{http://dx.doi.org/10.1103/PhysRevD.78.115003}{{\em
  Phys. Rev.} {\bfseries D78} (2008) 115003},
\href{http://arxiv.org/abs/0807.4668}{{\ttfamily arXiv:0807.4668 [hep-ph]}}.

\bibitem{Baro:2009gn}
N.~Baro and F.~Boudjema, ``{Automatised full one-loop renormalisation of the
  MSSM II: The chargino-neutralino sector, the sfermion sector and some
  applications},'' \href{http://dx.doi.org/10.1103/PhysRevD.80.076010}{{\em
  Phys.Rev.} {\bfseries D80} (2009) 076010},
\href{http://arxiv.org/abs/0906.1665}{{\ttfamily arXiv:0906.1665 [hep-ph]}}.

\bibitem{Baro:2007em}
N.~Baro, F.~Boudjema, and A.~Semenov, ``{Full one-loop corrections to the relic
  density in the MSSM: A Few examples},''
  \href{http://dx.doi.org/10.1016/j.physletb.2008.01.031}{{\em Phys. Lett.}
  {\bfseries B660} (2008) 550--560},
\href{http://arxiv.org/abs/0710.1821}{{\ttfamily arXiv:0710.1821 [hep-ph]}}.

\bibitem{Baro:2009ip}
N.~Baro, G.~Chalons, and S.~Hao, ``{Coannihilation with a chargino and gauge
  boson pair production at one-loop},''
  \href{http://dx.doi.org/10.1063/1.3327540}{{\em AIP Conf. Proc.} {\bfseries
  1200} (2010) 1067--1070},
\href{http://arxiv.org/abs/0909.3263}{{\ttfamily arXiv:0909.3263 [hep-ph]}}.

\bibitem{Baro:2009na}
N.~Baro, F.~Boudjema, G.~Chalons, and S.~Hao, ``{Relic density at one-loop with
  gauge boson pair production},''
  \href{http://dx.doi.org/10.1103/PhysRevD.81.015005}{{\em Phys. Rev.}
  {\bfseries D81} (2010) 015005},
\href{http://arxiv.org/abs/0910.3293}{{\ttfamily arXiv:0910.3293 [hep-ph]}}.

\bibitem{Semenov:1998eb}
A.~Semenov, ``{LanHEP: A package for automatic generation of Feynman rules from
  the Lagrangian},''
\href{http://dx.doi.org/10.1016/S0010-4655(98)00143-X}{{\em Comput. Phys.
  Commun.} {\bfseries 115} (1998) 124--139}.

\bibitem{Semenov:2008jy}
A.~Semenov, ``{LanHEP: A Package for the automatic generation of Feynman rules
  in field theory. Version 3.0},''
  \href{http://dx.doi.org/10.1016/j.cpc.2008.10.012}{{\em Comput. Phys.
  Commun.} {\bfseries 180} (2009) 431--454},
\href{http://arxiv.org/abs/0805.0555}{{\ttfamily arXiv:0805.0555 [hep-ph]}}.

\bibitem{Semenov:2014rea}
A.~Semenov, ``{LanHEP - a package for automatic generation of Feynman rules
  from the Lagrangian. Updated version 3.2},''
\href{http://arxiv.org/abs/1412.5016}{{\ttfamily arXiv:1412.5016
  [physics.comp-ph]}}.

\bibitem{Hahn:2000kx}
T.~Hahn, ``{Generating Feynman diagrams and amplitudes with FeynArts 3},''
  \href{http://dx.doi.org/10.1016/S0010-4655(01)00290-9}{{\em Comput. Phys.
  Commun.} {\bfseries 140} (2001) 418--431},
\href{http://arxiv.org/abs/hep-ph/0012260}{{\ttfamily arXiv:hep-ph/0012260
  [hep-ph]}}.

\bibitem{Hahn:1998yk}
T.~Hahn and M.~Perez-Victoria, ``{Automatized one loop calculations in
  four-dimensions and D-dimensions},''
  \href{http://dx.doi.org/10.1016/S0010-4655(98)00173-8}{{\em Comput. Phys.
  Commun.} {\bfseries 118} (1999) 153--165},
\href{http://arxiv.org/abs/hep-ph/9807565}{{\ttfamily arXiv:hep-ph/9807565
  [hep-ph]}}.

\bibitem{Hahn:2000jm}
T.~Hahn, ``{Automatic loop calculations with FeynArts, FormCalc, and
  LoopTools},'' \href{http://dx.doi.org/10.1016/S0920-5632(00)00848-3}{{\em
  Nucl. Phys. Proc. Suppl.} {\bfseries 89} (2000) 231--236},
\href{http://arxiv.org/abs/hep-ph/0005029}{{\ttfamily arXiv:hep-ph/0005029
  [hep-ph]}}.

\bibitem{Chalons:2011ia}
G.~Chalons and A.~Semenov, ``{Loop-induced photon spectral lines from
  neutralino annihilation in the NMSSM},''
  \href{http://dx.doi.org/10.1007/JHEP12(2011)055}{{\em JHEP} {\bfseries 12}
  (2011) 055},
\href{http://arxiv.org/abs/1110.2064}{{\ttfamily arXiv:1110.2064 [hep-ph]}}.

\bibitem{Chalons:2012xf}
G.~Chalons, M.~J. Dolan, and C.~McCabe, ``{Neutralino dark matter and the Fermi
  gamma-ray lines},''
  \href{http://dx.doi.org/10.1088/1475-7516/2013/02/016}{{\em JCAP} {\bfseries
  1302} (2013) 016},
\href{http://arxiv.org/abs/1211.5154}{{\ttfamily arXiv:1211.5154 [hep-ph]}}.

\bibitem{Chalons:2012qe}
G.~Chalons and F.~Domingo, ``{Analysis of the Higgs potentials for two doublets
  and a singlet},'' \href{http://dx.doi.org/10.1103/PhysRevD.86.115024}{{\em
  Phys. Rev.} {\bfseries D86} (2012) 115024},
\href{http://arxiv.org/abs/1209.6235}{{\ttfamily arXiv:1209.6235 [hep-ph]}}.

\bibitem{Belanger:2014roa}
G.~Belanger, V.~Bizouard, and G.~Chalons, ``{Boosting Higgs boson decays into
  gamma and a Z in the NMSSM},''
  \href{http://dx.doi.org/10.1103/PhysRevD.89.095023}{{\em Phys. Rev.}
  {\bfseries D89} no.~9, (2014) 095023},
\href{http://arxiv.org/abs/1402.3522}{{\ttfamily arXiv:1402.3522 [hep-ph]}}.

\bibitem{Belanger:2003sd}
G.~Belanger, F.~Boudjema, J.~Fujimoto, T.~Ishikawa, T.~Kaneko, K.~Kato, and
  Y.~Shimizu, ``{Automatic calculations in high energy physics and Grace at
  one-loop},'' \href{http://dx.doi.org/10.1016/j.physrep.2006.02.001}{{\em
  Phys. Rept.} {\bfseries 430} (2006) 117--209},
\href{http://arxiv.org/abs/hep-ph/0308080}{{\ttfamily arXiv:hep-ph/0308080
  [hep-ph]}}.

\bibitem{Passarino:1978jh}
G.~Passarino and M.~J.~G. Veltman, ``{One Loop Corrections for e+ e-
  Annihilation Into mu+ mu- in the Weinberg Model},''
\href{http://dx.doi.org/10.1016/0550-3213(79)90234-7}{{\em Nucl. Phys.}
  {\bfseries B160} (1979) 151--207}.

\bibitem{Chankowski:1992er}
P.~H. Chankowski, S.~Pokorski, and J.~Rosiek, ``{Complete on-shell
  renormalization scheme for the minimal supersymmetric Higgs sector},''
  \href{http://dx.doi.org/10.1016/0550-3213(94)90141-4}{{\em Nucl. Phys.}
  {\bfseries B423} (1994) 437--496},
\href{http://arxiv.org/abs/hep-ph/9303309}{{\ttfamily arXiv:hep-ph/9303309
  [hep-ph]}}.

\bibitem{Dabelstein:1994hb}
A.~Dabelstein, ``{The One loop renormalization of the MSSM Higgs sector and its
  application to the neutral scalar Higgs masses},''
  \href{http://dx.doi.org/10.1007/BF01624592}{{\em Z. Phys.} {\bfseries C67}
  (1995) 495--512},
\href{http://arxiv.org/abs/hep-ph/9409375}{{\ttfamily arXiv:hep-ph/9409375
  [hep-ph]}}.

\bibitem{bizouard:tel-01447488}
V.~Bizouard, {\em {Precision calculations in the Next-to-Minimal Supersymmetric
  Standard Model}}.
\newblock Theses, {Universit{\'e} Grenoble Alpes}, Oct., 2015.
\newblock
  \href{http://arxiv.org/abs/https://hal.archives-ouvertes.fr/tel-01447488}{{\%
ttfamily https://hal.archives-ouvertes.fr/tel-01447488}}.

\end{thebibliography}

\providecommand{\href}[2]{#2}\begingroup\raggedright\endgroup

\end{document}